\documentclass[aps,english,prd,floatfix,amsmath,eqsecnum, twocolumn, superscriptaddress, longbibliography]{revtex4-1}

\usepackage{graphicx,color,framed}
\usepackage{hyperref}
\usepackage{times}
\usepackage{enumerate}
\usepackage{lipsum}
\usepackage{slashed}
\usepackage{array}
\usepackage{textcomp}
\usepackage{amsthm}
\usepackage{amssymb}
\usepackage{comment}
\usepackage{bm}
\usepackage{bbm}
\usepackage{tikz}
\usepackage{pgfplots}
\usepackage{makecell}
\usepackage[normalem]{ulem}
\pgfplotsset{compat = newest}

\newcommand{\bpm}{\begin{pmatrix}}
\newcommand{\epm}{\end{pmatrix}}
\newcommand{\D}[1]{\text{d}#1}

\newcommand{\tr}{\,{\rm Tr}\,}

\newcommand{\<}{\langle}
\renewcommand{\>}{\rangle}
\renewcommand{\vec}[1]{{\bf #1}}

\hypersetup{
    colorlinks=true, 
    linktoc=all,     
    linkcolor=blue,  
}

\begin{document}

\title{Effect of slowly decaying long-range interactions on topological qubits}

\author{Etienne Granet}
\thanks{The work was performed at the Kadanoff Center for Theoretical Physics at the University of Chicago.}
\affiliation{Quantinuum, Leopoldstrasse 180, 80804 Munich, Germany}
\author{Michael Levin}
\affiliation{Leinweber Institute for Theoretical Physics, University of Chicago, Chicago, Illinois 60637,  USA}

\begin{abstract}
We study the robustness of topological ground state degeneracy to long-range interactions in quantum many-body systems. We focus on slowly decaying two-body interactions that scale like a power-law $1/r^\alpha$ where $\alpha$ is smaller than the spatial dimension; such interactions are beyond the reach of known stability theorems which only apply to short-range or rapidly decaying long-range perturbations. Our main result is a computation of the ground state splitting of several toy models, which are variants of the 1D Ising model $H = -\sum_i \sigma^z_i \sigma^z_{i+1} + \lambda \sum_{ij} |i-j|^{-\alpha} \sigma^x_i \sigma^x_j$ with $\lambda > 0$ and $\alpha < 1$. In one variant, the power-law interactions are replaced by \emph{all-to-all} interactions, $\frac{\lambda}{4 L^\alpha}\sum_{ij} \sigma^x_i \sigma^x_j$, where $L$ is the system size, while the other variant has true power-law interactions but is built out of quantum rotors rather than Ising spins.
These models are also closely connected to the Kitaev p-wave wire model with power-law density-density interactions. In these examples, we find that the splitting $\delta$ scales like a \emph{stretched} exponential $\delta \sim \exp(-C L^{\frac{1+\alpha}{2}})$. 
 Our computations are based on path integral techniques similar to the instanton method introduced by Coleman. We also study another toy model with long-range interactions that can be analyzed without path integral techniques and that shows similar behavior.
\end{abstract}

\maketitle

\section{Introduction}

A ``topological qubit'' is a gapped quantum many-body system with two special properties: (i) the system has multiple nearly degenerate ground states with an energy splitting that is exponentially small in the system size, and (ii) this exponentially small ground state splitting is stable under small perturbations of the Hamiltonian~\cite{wen-niu, kitaev1997fault, kitaev-wire}. Systems of this kind are attractive for quantum computing applications because their ground state subspace can store quantum information that is intrinsically protected against errors and decoherence~\cite{nayak2008non}.

The most well-understood examples of topological qubits involve lattice models with \emph{short-range} interactions like the toric code model. For many of these models one can rigorously prove that their ground state splitting is exponentially small in the presence of arbitrary short-range perturbations of the Hamiltonian~\cite{kirkwood1983expansions,klich2010on,bravyi2010topological,bravyi2011a,michalakis2013stability,nachtergaele2019quasi,nachtergaele2019quasibis}.  

On the other hand, we are lacking a similar level of understanding of topological qubits with \emph{long-range} (e.g.~power-law) interactions. To see the problem, imagine perturbing one of the prototypical lattice models for topological qubits by adding a weak power-law interaction. Assume that the perturbation is small enough that the gap does not close. The standard way to bound the ground state splitting in this situation is to use the ``quasi-adiabatic continuation approach'' of Ref.~\onlinecite{QAC}. However, if the Hamiltonian includes power-law interactions, this approach gives a bound on the splitting that decays like a \emph{power-law} in the system size~\cite{michalakis2013stability} rather than an exponential.\footnote{This slower decay is due to the fact that the unitary transformation associated with quasi-adiabatic continuation is not as local as it is in the short-range case: it transforms local operators into quasi-local operators with tails that decay like a power-law or slower.} The weakness of this bound raises the possibility that long-range interactions could fundamentally change the behavior of topological qubits, lifting their exponentially small ground state splitting. 

In this paper, we investigate this question in the context of a minimal toy model introduced by Ref.~\onlinecite{lapa21stability}. We will argue below that this model captures the basic physics of a topological qubit with long-range interactions. Specifically, we consider a spin-$1/2$ chain made up of $L$ spins with periodic boundary conditions. The Hamiltonian is given by
\begin{equation}
    H = H_0  + \lambda V,
\label{H}
\end{equation}
where $H_0$ is the Ising Hamiltonian
\begin{align}
H_0=-\sum_{j=1}^L \sigma_j^z \sigma^z_{j+1},
\label{H0}
\end{align}
and $V$ is a long-range spin-spin interaction of the form    
\begin{equation}    
    V=\sum_{i,j=1}^L f(|i-j|)\sigma^x_i \sigma^x_j
    \label{V}
\end{equation}
Here, $\lambda$ is a real parameter and $f(r)$ is a dimensionless real function. We will be particularly interested in the case where $f(r)$ has a power-law behavior, but we will not impose any restrictions on $f(r)$ for now, except for the normalization condition $|f(r)| \leq 1$ and the requirement $f(r) = f(L-r)$, which is needed for consistency with periodic boundary conditions. 

We can think of $H$ as a close analog of a traditional topological qubit with long-range interactions. To see the analogy, notice that the unperturbed model $H_0$ has two degenerate ground states. This degeneracy originates from the fact that $H_0$ spontaneously breaks the Ising symmetry
\begin{equation}
    S=\prod_{j=1}^L \sigma_j^x\,.
    \label{Sising}
\end{equation}
Furthermore this ground state degeneracy is \emph{almost} as robust as that of a topological qubit: the ground state splitting of $H_0$ is known to be exponentially small in the system size $L$ in the presence of arbitrary Ising \emph{symmetric} short-range perturbations~\cite{bravyi2011a}. In this sense, $H_0$ can be thought of as a ``symmetry-protected'' analog of a traditional topological qubit like the toric code model. Likewise, we can view the full Hamiltonian $H$ as analogous to a topological qubit with a long-range two-body interaction.

In fact, the connection between our spin model $H$ and traditional topological qubits is more than just an analogy: using a Jordan-Wigner transformation, we can map the spin Hamiltonian $H$ (\ref{H}) (albeit with open boundary conditions) onto the following fermionic Hamiltonian (also with open boundary conditions):
\begin{align}
    H_f &=-\sum_{j=1}^{L-1} (c^\dagger_j c_{j+1}+c_j c_{j+1}+h.c.) \nonumber \\
    &+ 4 \lambda\sum_{i,j=1}^L  f(|i-j|) \left(n_i - \frac{1}{2} \right) \left (n_j - \frac{1}{2} \right)\,,
\end{align}
This is precisely the Kitaev p-wave wire Hamiltonian~\cite{kitaev-wire}, perturbed by long-range density-density interactions. 

Having motivated the spin Hamiltonian $H$ as a reasonable toy model for a topological qubit with long-range interactions, we now return to our original question: understanding the behavior of the ground state splitting of $H$ once we turn on a small $\lambda \neq 0$. More precisely, the key question is whether the ground state splitting is \emph{exponentially} small in system size $L$, as it would be if $V$ were an (Ising symmetric) short-range interaction. 

This question was partially answered by Ref.~\onlinecite{lapa21stability}, which showed that the ground state splitting is exponentially small for any long-range interaction $f(r)$ that satisfies
\begin{equation}\label{summability0}
    \sum_{r=0}^{L-1} |f(r)| \leq c\,,
\end{equation}
for some constant $c$ that can be chosen to be independent of the system size $L$. More precisely, Ref.~\onlinecite{lapa21stability} showed that if $f(r)$ obeys (\ref{summability0}), then there exists some constant $\lambda_0 > 0$ such that if $|\lambda| < \lambda_0$, the splitting $\delta$ can be upper bounded as
\begin{align}
|\delta| \leq c_1 L e^{-c_2 L},
\end{align}
for some constants $c_1, c_2 > 0$. Here, $\delta$ is defined by 
\begin{align}
\delta = E_- - E_+
\label{deltadef}
\end{align}
where $E_\pm$ denote the ground state energies of $H$ in the $S = \pm 1$ sectors. The proof of this result was based on a perturbative expansion in $\lambda$, which can be shown to be convergent as long as $f(r)$ obeys (\ref{summability0}). We will refer to (\ref{summability0}) as the  ``summability condition.''

More recently, Ref.~\onlinecite{YinLucas2025} derived a similar exponential bound on the ground state splitting for a large class of stabilizer Hamiltonians (e.g.~the toric code model) perturbed by weak long-range interactions. Similarly to Ref.~\onlinecite{lapa21stability}, this bound holds for perturbations that obey a summability condition like (\ref{summability0}). (See also Refs. \onlinecite{lavasani2024stability, de2024ldpc} for related stability results for quantum low density parity check codes).

These results represent significant progress but they have a limitation: they don't tell us anything about the effects of long-range interactions that violate the summability condition (\ref{summability0}). This is an important issue since non-summable long-range interactions occur in many physical systems of interest. For example, consider (two-dimensional) fractional quantum Hall systems with Couloumb interactions: in this case the analog of the sum in Eq.~(\ref{summability0}) is divergent since $f(r) \sim 1/r$. Given this behavior, one might worry that the physics of fractional quantum Hall topological qubits~\cite{wen-niu} is more similar to the non-summable case where the behavior of the splitting is not understood. 

Motivated by these concerns, the goal of this paper is to understand the case where the summability condition (\ref{summability0}) is \emph{not} satisfied. More specifically, we consider power-law interactions $f(r) \sim 1/r^\alpha$ with exponent $0 < \alpha < 1$.\footnote{Actually, due to our periodic boundary conditions, we will consider $f(r)$ of the form $f(r) \sim 1/ \min(r,L-r)^\alpha$.} We also assume that $\lambda > 0$ since this is the most interesting and physically relevant scenario: when $\lambda$ is negative, a simple variational argument shows that $H$ has an \emph{instability} at infinitesimal $\lambda$ to a paramagnetic state without any ground state degeneracy~\cite{lapa21stability}.

Unfortunately we are not able to directly compute the ground state splitting of $H$ for the case $f(r) \sim 1/r^\alpha$. Instead, we compute $\delta$ for two variants of this model. The first variant consists of the Hamiltonian $H$ (\ref{H}) but with $f(r) = 1/(4L^\alpha)$. In other words, instead of power-law interactions, we consider \emph{all-to-all} interactions that scale with the system size like a power-law. The second variant (\ref{confinedh}) has true power-law interactions $f(r) \sim 1/r^\alpha$ but is built out of quantum rotors rather than Ising spins. For both of these variants, we find that the splitting $\delta$ scales like a \emph{stretched} exponential in the system size $L$, namely
\begin{align}
    \log \delta \sim -(\text{const.}) L^{\frac{1 +\alpha}{2}}
\label{delta-stretched}
\end{align}
We conjecture that the above stretched exponential scaling of the splitting also occurs in the original Ising spin chain $H$ (\ref{H}) with $f(r) \sim 1/r^\alpha$, though we don't have a well-controlled calculation in this case. For both of these models, we compute $\delta$ using path integral techniques similar to the instanton method introduced by Coleman~\cite{coleman1988aspects}. 

We also discuss a third variant of $H$ where our calculation of $\delta$ does not require any path integral techniques. Specifically, we consider a model (\ref{toymodel}) where $H_0$ is replaced by a projector onto the two-dimensional ground state of $-\sum_{j=1}^L \sigma_j^z\sigma_{j+1}^z$, and where $V$ is replaced by an all-to-all interaction of the form $\frac{1}{L^\alpha}\left(\sum_j \mathcal{O}_j \right)^2$ with $\mathcal{O}_j$ being a general, Ising symmetric, local Hermitian operator. While this model is non-local, it has the advantage that the splitting $\delta$ can be computed explicitly for general choices of the operators $\mathcal{O}_j$. For this model, we again find stretched exponential scaling (\ref{delta-stretched}) for several choices of $\mathcal{O}_j$, but we also find that the scaling of the splitting can be sensitive to the detailed structure of the operators $\mathcal{O}_j$. It is unclear whether this sensitivity to details is specific to this (non-local) model or whether it is a more general feature of topological qubits with long-range interactions. 

This paper is organized as follows: First, in Sec.~\ref{sec:all-to-all}, we study the Ising spin chain (\ref{H}) with all-to-all interactions, $f(r) = 1/(4L^\alpha)$ with $\alpha < 1$. We start with this all-to-all spin chain because it is the simplest of our three toy models, and at the same time it captures the basic physics of slowly decaying power-law interactions. Then, in  
Sec.~\ref{sec:rotor}, we study a quantum rotor model (\ref{confinedh}) which is analogous to the Ising spin chain and has true power-law interactions $f(r) \sim 1/r^\alpha$. Next, in Sec.~\ref{sec:toy}, we present the non-local model (\ref{toymodel}) where the splitting can be computed explicitly for several different types of all-to-all interactions. Finally, in Sec.~\ref{sec:disc} we summarize our results and discuss future directions.

\section{Ising spin chain with all-to-all interactions\label{sec:all-to-all}}

\subsection{Hamiltonian}
In this section, we consider the Ising spin chain (\ref{H}) in the case where $f(r) = 1/(4L^\alpha)$. This choice of $f(r)$ can be thought of as a toy version of power-law interactions: it corresponds to \emph{all-to-all} interactions that scale with the system size like a power-law. With this choice of $f(r)$, the Hamiltonian takes the form
\begin{equation}\label{ising}
H=-\sum_{j=1}^L \sigma^z_j\sigma^z_{j+1}+\frac{\lambda}{4L^\alpha} \left(\sum_{j=1}^L\sigma^x_j\right)^2\,,
\end{equation}
We assume that $\lambda > 0$ and $0 < \alpha < 1$, as this is the regime where our path integral approach is well-controlled. We also assume that $L$ is a multiple of $4$ for simplicity.\footnote{It is natural to assume that $L$ is even since $H$ has an exact ground state degeneracy for odd $L$ due to Kramers theorem. The reason we require $L$ to be a multiple of $4$ is for convenience: this choice avoids factors of $(-1)^{L/2}$ which lead to an alternating sign of the splitting $\delta$ (\ref{deltadef}).}

We note that this kind of all-to-all interaction typically arises in models when the interaction of the spins with an external mode is integrated out. For example it arises in a particle-number conserving version of the Kitaev model~\cite{lapa2020rigorous} or in variants of the Dicke model of quantum optics~\cite{rohn2020ising} (though $\lambda < 0$ in the latter case). 

Let $\delta$ denote the ground state splitting between the $S = \pm 1$ sectors, where $S$ denotes the Ising symmetry transformation (\ref{Sising}). Our main result is that $\delta$ scales as the stretched exponential
\begin{align}
    \log \delta = -\frac{2}{\sqrt{\lambda}} L^{\frac{1+\alpha}{2}}\,,
    \label{deltaformalltoall}
\end{align}
in the limit of large $L$. Here, the above formula should be thought of as the leading order term in an asymptotic series for $\log \delta$; in particular, we expect that there are subleading corrections to this formula of order $L^\alpha$.

\subsection{Preliminaries}

Our basic strategy for deriving (\ref{deltaformalltoall}) is to first consider the finite temperature analog of $\delta$, namely the \emph{free} energy difference between the two sectors. We denote this free energy difference by $\delta F_\beta$ where $\beta$ is the inverse temperature. Formally, $\delta F_\beta$ is defined by:
\begin{equation}
    \delta F_\beta=\frac{1}{\beta}\log\tr[e^{-\beta H}\tfrac{1+S}{2}]-\frac{1}{\beta}\log\tr[e^{-\beta H}\tfrac{1-S}{2}]\,.
\end{equation}
In what follows, we will focus on computing the free energy difference $\delta F_\beta$ for large $L$. Once we complete this calculation, we will then derive the ground state splitting $\delta$ using the fact that $\delta F_\beta \rightarrow \delta$ as $\beta \rightarrow \infty$.

Following this program, we first rewrite $\delta F_\beta$ as
\begin{equation}
    \delta F_\beta=\frac{1}{\beta}\log \frac{1+s_\beta}{1-s_\beta}\,,
\label{deltasplit}
\end{equation}
with
\begin{equation}
s_\beta=\frac{\tr[e^{-\beta H}S]}{\tr[e^{-\beta H}]}\,.
\end{equation}
Our task is now to compute $s_\beta$. We will accomplish this using a path integral formula.

\subsection{Path integral expression for $s_\beta$ \label{pathintegral}}
To derive the path integral formula for $s_\beta$, we write the Hamiltonian as a sum
\begin{equation}
    H=H_0+\frac{\lambda}{L^\alpha} (M^{x})^2\,,
\end{equation}
where $H_0$ is the unperturbed Hamiltonian and $M^x$ is the magnetization:
\begin{equation}\label{unperturbed}
    H_0=-\sum_{j=1}^L \sigma_j^z\sigma_{j+1}^z, \quad \quad \quad M^x=\frac{1}{2}\sum_{j=1}^L \sigma_j^x\,. 
\end{equation}
We then rewrite $s_\beta$ using the Trotter expansion:
\begin{align}
    s_\beta &= \frac{\tr \left[e^{-\beta\left(H_0 + \tfrac{\lambda}{L^\alpha} (M^{x})^2\right)} S\right]}{\tr \left[e^{-\beta\left(H_0 + \tfrac{\lambda}{L^\alpha} (M^{x})^2\right)} \right]} \nonumber \\
    &=
    \underset{m\to\infty}{\lim}\frac{\tr \left[\left(e^{-\tfrac{\beta}{m}H_0} e^{-\tfrac{\beta}{m}\tfrac{\lambda}{L^\alpha} (M^{x})^2} \right)^m S\right]}{\tr \left[\left(e^{-\tfrac{\beta}{m}H_0} e^{-\tfrac{\beta}{m}\tfrac{\lambda}{L^\alpha} (M^{x})^2} \right)^m \right]}\,.
    \label{eq:trotter}
\end{align}
Next, we use the Hubbard-Stratonovich identity to express $e^{-\tfrac{\beta}{m}\tfrac{\lambda}{L^\alpha} (M^{x})^2}$ as:
\begin{equation}
    e^{- \tfrac{\beta}{m}\tfrac{\lambda}{L^\alpha} (M^{x})^2}=\sqrt{\frac{\beta L^\alpha}{4\pi m\lambda}}\int_{-\infty}^{\infty} e^{- \tfrac{\beta}{m} (iM^x\varphi + \tfrac{L^\alpha \varphi^2}{4\lambda})}\D{\varphi}\,.
    \label{eq:hubbstrat}
\end{equation}
Substituting the integral (\ref{eq:hubbstrat}) into (\ref{eq:trotter}) in place of each $e^{-\tfrac{\beta}{m}\tfrac{\lambda}{L^\alpha} (M^{x})^2}$ term gives an $m$-dimensional integral,
\begin{widetext}
\begin{equation}
s_\beta = \underset{m\to\infty}{\lim} \frac{\int \prod_{i=1}^m d\varphi_i \tr \left[ \prod_{i=1}^m \left(e^{-\tfrac{\beta}{m}H_0} e^{-\tfrac{\beta}{m}(i M^x \varphi_i + \frac{L^\alpha \varphi_i^2}{4\lambda})}\right) S\right]} 
{\int \prod_{i=1}^m d\varphi_i \tr \left[ \prod_{i=1}^m \left(e^{-\tfrac{\beta}{m}H_0} e^{-\tfrac{\beta}{m}(i M^x \varphi_i + \frac{L^\alpha \varphi_i^2}{4\lambda})}\right) \right]}. 
\end{equation}
\end{widetext}
Taking the limit $m \rightarrow \infty$, gives a ratio of two path integrals
\begin{equation}
    s_\beta =\frac{\int \mathcal{D}[\varphi]\exp\left(-\mathcal{S}_n[\varphi]-\int_{0}^{\beta} \D{\tau} \tfrac{L^\alpha \varphi(\tau)^2}{4\lambda} \right)}{\int \mathcal{D}[\varphi]\exp\left(-\mathcal{S}_d[\varphi]-\int_{0}^{\beta} \D{\tau} \tfrac{L^\alpha \varphi(\tau)^2 }{4\lambda} \right)}\,,
\end{equation}
where the two path integrals run over arbitrary paths $\varphi(\tau)$ without any boundary conditions and where the two actions $\mathcal{S}_{n,d}[\varphi]$ in the numerator and denominator are defined as follows
\begin{align}
    e^{-\mathcal{S}_n[\varphi]}&=\tr \left[\mathcal{T}\exp \left(-\int_{0}^{\beta} \D{\tau}(H_0+iM^x\varphi(\tau)) \right)S \right] \nonumber \\
e^{-\mathcal{S}_d[\varphi]}&=\tr \left[\mathcal{T}\exp \left(-\int_{0}^{\beta} \D{\tau}(H_0+iM^x\varphi(\tau)) \right)\right]\,.
\label{SnSdphi}
\end{align}
To proceed further, we make a change of variable
\begin{equation}
    \theta(\tau)=\int_{0}^\tau \varphi(\tau') \D{\tau'}\,.
    \label{thetadef}
\end{equation}
Note that this definition implies that $\theta$ has the boundary condition $\theta(0)=0$. Our formula then becomes
\begin{align}
 s_\beta =\frac{\int \mathcal{D}[\theta]\exp\left(-\mathcal{S}^{\rm tot}_{n}[\theta] \right)}
{\int \mathcal{D}[\theta]\exp\left(-\mathcal{S}^{\rm tot}_{d}[\theta] \right)}\,, 
 \label{sbetaform}
\end{align}
where $\mathcal{S}^{\rm tot}_{n}[\theta]$ and $\mathcal{S}^{\rm tot}_{d}[\theta]$
are defined by
\begin{equation}\label{actiona}
    \mathcal{S}^{\rm tot}_{n,d}[\theta]=\mathcal{S}_{n,d}[\theta]+\frac{L^\alpha}{4\lambda}\int_0^\beta \D{\tau} \dot{\theta}(\tau)^2\,,
\end{equation}
with
\begin{align*}
    e^{-\mathcal{S}_n[\theta]}&=\tr \left[\mathcal{T}\exp \left(-\int_{0}^{\beta} \D{\tau}(H_0+iM^x\dot{\theta}(\tau)) \right)S \right] \nonumber \\
e^{-\mathcal{S}_d[\theta]}&=\tr \left[\mathcal{T}\exp \left(-\int_{0}^{\beta} \D{\tau}(H_0+iM^x\dot{\theta}(\tau)) \right)\right]\,.
\end{align*}
Finally, we rewrite the two actions $\mathcal{S}_n, \mathcal{S}_d$ in a more convenient form using the following operator identity, which applies to arbitrary one-parameter families of operators $A(\tau), B(\tau)$ depending smoothly on a real variable $\tau$, such that $[\dot{B}(\tau), B(\tau')] = 0$ for any $\tau,\tau '$:
\begin{align}\label{torder}
&\mathcal{T}\exp\left(\int_{0}^{\beta} \D{\tau}(A(\tau)+\dot{B}(\tau)) \right)= \nonumber \\
&e^{B(\beta)}\mathcal{T}\exp\left(\int_{0}^{\beta} \D{\tau}e^{-B(\tau)}A(\tau) e^{B(\tau)} \right)e^{-B(0)} \,.
\end{align}
(To derive this identity, define $X(\tau) = e^{-B(\tau)} \mathcal{T}\exp\left(\int_{0}^{\tau} \D{\tau'}(A(\tau')+\dot{B}(\tau')) \right)$ and note that $X(\tau)$ obeys the differential equation $\frac{d}{d\tau} X(\tau) = (e^{-B(\tau)}A(\tau) e^{B(\tau)}) X(\tau)$, as well as $X(0) = e^{-B(0)}$). Using (\ref{torder}) with $A = -H_0$ and $B(\tau) = -i M^x \theta(\tau)$, we obtain the expressions
\begin{align}
    e^{-\mathcal{S}_n[\theta]}&=\tr \left[\mathcal{T}\exp \left(-\int_{0}^{\beta} \D{\tau}W(\theta(\tau)) \right)S e^{-iM^x\theta(\beta)} \right] \label{expsaz}\\
e^{-\mathcal{S}_d[\theta]}&=\tr \left[\mathcal{T}\exp \left(-\int_{0}^{\beta} \D{\tau}W(\theta(\tau)) \right)e^{-iM^x\theta(\beta)}\right]\,,
\label{SnSd}
\end{align}
where $W(\theta)$ is defined by
\begin{equation}\label{w}
    W(\theta)=e^{iM^x\theta}H_0e^{-iM^x\theta}\,.
\end{equation}

\subsection{Saddle point approximation\label{saddlesection}}

We now determine the leading order behavior of $s_\beta$ at large $L$. The key idea behind our calculation is that when $L$ is large, the path integrals in the numerator and denominator of (\ref{sbetaform}) are each dominated by a single path, or a small family of paths, together with small fluctuations about these paths. To see why, note that the actions $\mathcal{S}_n[\theta], \mathcal{S}_d[\theta]$ are expected to scale \emph{extensively} with $L$ for large $L$: that is, they should take the form
\begin{align}
\mathcal{S}_{n,d}[\theta] = L f_{n,d}[\theta], 
\end{align}
for some functionals $f_n[\theta], f_d[\theta]$. Also, we can see that the kinetic term $\dot{\theta}^2$ in $\mathcal{S}^{\rm tot}_{n}[\theta]$ and $\mathcal{S}^{\rm tot}_{d}[\theta]$ has a coefficient of order $L^\alpha$. Therefore, every term in $\mathcal{S}^{\rm tot}_{n}[\theta], \mathcal{S}^{\rm tot}_{d}[\theta]$ has a coefficient that diverges as $L \rightarrow \infty$, so fluctuations are indeed suppressed in this limit.

This suppression of fluctuations means that we can get a good approximation to the two path integrals (and hence $s_\beta$) using a saddle point approximation. We now carry out this saddle point analysis.

\subsubsection{Saddle point for $\mathcal{S}^{\rm tot}_{d}[\theta]$}
Our first task is to find the dominant saddle points -- i.e. find the paths $\bar{\theta}_n(\tau)$ and $\bar{\theta}_d(\tau)$ that control the path integrals in the numerator and denominator of (\ref{sbetaform}). We start with the denominator of (\ref{sbetaform}). In this case, the dominant saddle point $\bar{\theta}_d(\tau)$ is very simple:
\begin{align}
    \bar{\theta}_d(\tau) = 0, \quad \quad 0 \leq \tau \leq \beta\,.
\end{align}
The reason that this path is the dominant saddle point is that it has two important properties: (i) it is the (unique) global minimum of $\Re(\mathcal{S}^{\rm tot}_{d}[\theta])$, and (ii) it obeys the saddle point condition $\frac{\delta \mathcal{S}^{\rm tot}_{d}}{\delta \theta} = 0$. Together these two properties imply that the path integral in the denominator is dominated by the path $\bar{\theta}_d(\tau)$ together with small fluctuations around this path.

We now derive the above properties (i)-(ii). To derive property (i), note that $\bar{\theta}_d(\tau)$ clearly minimizes the kinetic term in $\Re(\mathcal{S}^{\rm tot}_{d}[\theta])$, since this term is proportional to $\dot{\theta}^2$. Furthermore, $\bar{\theta}_d(\tau)$ also minimizes the other term in the action, $\Re(\mathcal{S}_d[\theta])$: we prove this result in Appendix~\ref{holder}
using the H\"older inequality for matrices. As for property (ii), this follows from an explicit calculation: from (\ref{SnSd}) we see that the variation $\delta \mathcal{S}_d$ about the path $\bar{\theta}_d$ is given by
\begin{align}
    \delta \mathcal{S}_d = \int_0^\beta d\tau \ \< W'(0)\>_\beta \cdot \delta \theta(\tau) + i \<M^x\>_\beta \cdot \delta \theta(\beta)
\end{align}
where $\<\cdot\>_\beta$ is the expectation value in the Gibbs state for $H_0$ at inverse temperature $\beta$. Next observe that $\<M^x\>_\beta = 0$, and $\<W'(0)\>_\beta = 0$ as well, since $W'(0) = i [M^x, H_0]$. Hence $\frac{\delta \mathcal{S}_d}{\delta \theta} = 0$. Finally, since the kinetic term also has a vanishing variation about the path $\bar{\theta}_d$, we conclude that $\frac{\delta \mathcal{S}^{\rm tot}_{d}}{\delta \theta} = 0$.

\subsubsection{Saddle point for $\mathcal{S}^{\rm tot}_{n}[\theta]$}
\label{minimalpath}
The dominant saddle point for the numerator of (\ref{sbetaform}) is less obvious. To find this saddle point, we use the following strategy. First, we search for a path $\bar{\theta}_n$ that minimizes $\Re(\mathcal{S}^{\rm tot}_{n}[\theta])$. We then show that this minimal path $\bar{\theta}_n$ obeys the saddle point condition $\frac{\delta \mathcal{S}^{\rm tot}_{n}}{\delta \theta} = 0$. It then follows immediately that $\bar{\theta}_n$  (together with small fluctuations) dominates the path integral. 

To find the minimal path $\bar{\theta}_n$, we need to make two assumptions about its basic structure: 
\begin{itemize}
\item{$\bar{\theta}_n$ obeys the boundary conditions
\begin{equation}
\bar{\theta}_n(0) = 0, \quad \quad \bar{\theta}_n(\beta) = \pi
\label{instantonbc}
\end{equation}}
\item{$\bar{\theta}_n$ is approximately a step function in the sense that
\begin{equation}
    \bar{\theta}_n(\tau) = \pi \Theta(\tau - \tau^*) + 
    \mathcal{O}(e^{-|\tau - \tau^*|/\Delta \tau})\,,
    \label{instantonstruct}
\end{equation}
for some $\tau^*$ and some time scale $\Delta \tau$ with $\Delta \tau \rightarrow 0$ as $L \rightarrow \infty$.} 
\end{itemize}

We give a detailed justification of these assumptions in Appendix~\ref{saddlepointapp}, but the basic intuition is as follows. To understand assumption (\ref{instantonbc}), notice that $S = e^{i \pi M^x}$.\footnote{ For general system sizes, $S = (-i)^L e^{i \pi M^x}$, but here we drop the factor of $(-i)^L$ since we are assuming $L$ is a multiple of $4$.} This means that the factor of $e^{-i M^x \theta(\beta)}$ in the expression for $\mathcal{S}_n[\theta]$ cancels the symmetry factor $S$ if and only if $\theta(\beta) = \pi \pmod{2 \pi}$. This explains why the minimal path obeys the boundary condition $\bar{\theta}_n(\beta) = \pi$: if this boundary condition is not satisfied, then there is a residual factor of $e^{i [\pi - \theta(\beta)] M^x}$ which suppresses the trace that defines $e^{- \mathcal{S}_n[\theta]}$. [Likewise, the boundary condition $\bar{\theta}_n(0) = 0$ follows from the definition (\ref{thetadef})]. As for the assumption (\ref{instantonstruct}), this comes from the fact that when $\theta$ changes with time, the operator $W(\theta)$, and hence its ground state, also changes with time. This time dependence suppresses the time ordered exponential in $e^{- \mathcal{S}_n[\theta]}$. As a result, the minimal path must have a $\theta$ that is constant everywhere except for a short period of time, leading to the step-like structure in (\ref{instantonstruct}). 

At this point, the reader may have noticed that $\bar{\theta}_n(\tau)$ has a similar step-like structure to the usual ``instanton'' that appears in quantum mechanical tunneling problems, see e.g. Ref.~\onlinecite{coleman1988aspects}. Indeed, we will see that $\bar{\theta}_n(\tau)$ is a close analogue of the usual instanton, and we will use this terminology below.

We now derive the precise functional form of $\bar{\theta}_n(\tau)$ starting from the assumptions (\ref{instantonbc}-\ref{instantonstruct}). The main challenge in doing so is that the action $\mathcal{S}^{\rm tot}_{n}[\theta]$  is not easy to work with, due to the time ordered exponential. Therefore, we will first derive a simpler ``reduced'' action $\mathcal{S}_{\rm reduced}[\theta]$ that approximately agrees with $\mathcal{S}^{\rm tot}_{n}[\theta]$ on paths with the step-like structure described in Eq.~(\ref{instantonstruct}). We will then find the desired path by minimizing $\Re(\mathcal{S}_{\rm reduced}[\theta])$.

To derive the reduced action, we use \eqref{torder}, with $A = H_0 - W(\theta(\tau))$ and $B = -\tau H_0 $, to write
\begin{align}
   &\mathcal{T}\exp \left(-\int_{0}^{\beta} \D{\tau}W(\theta(\tau)) \right)= \nonumber \\
   &e^{-\beta H_0} \mathcal{T}\exp \left(-\int_{0}^{\beta} \D{\tau}e^{\tau H_0}[W(\theta(\tau))-H_0]e^{-\tau H_0}
    \right)\,.
\end{align}
We then observe that since $W(0) = W(\pi) = H_0$, the integrand $W(\theta) - H_0$ is vanishingly small everywhere except near $\tau = \tau^*$ for any step-like path of the above type (\ref{instantonstruct}). Therefore we can make the following approximation to leading order in $\Delta \tau$:
\begin{align}
   &\mathcal{T}\exp \left(-\int_{0}^{\beta} \D{\tau}W(\theta(\tau)) \right) \approx \nonumber \\
   &e^{(\tau^*-\beta) H_0}\exp \left(-\int_{0}^{\beta} \D{\tau}[W(\theta(\tau))-H_0]
    \right)e^{-\tau^* H_0}\,.
\label{approx}
\end{align}
Next we substitute (\ref{approx}) into our expression for $\mathcal{S}_n[\theta]$ (\ref{expsaz}). Using the boundary condition $\theta(\beta) = \pi$ from (\ref{instantonbc}) together with the fact that $e^{i M^x \pi} = S$, we see that (\ref{expsaz}) simplifies to
\begin{equation}\label{esa}
    e^{-\mathcal{S}_n[\theta]}=\tr \left[e^{-\beta H_0}\exp \left(-\int_{0}^{\beta} \D{\tau}[W(\theta(\tau))-H_0]
    \right)\right]\,.
\end{equation}
Equivalently, we have
\begin{align}
     \mathcal{S}_n[\theta] &=-\log\tr[e^{-\beta H_0}] \nonumber \\
     &-\log \left\langle \exp \left(-\int_{0}^{\beta} \D{\tau}[W(\theta(\tau))-H_0]
    \right) \right\rangle_\beta\,,
    \label{esa2}
\end{align}
where $\langle \cdot \rangle_\beta$ denotes the expectation value in the Gibbs state for $H_0$ at inverse temperature $\beta$.

We now evaluate the expectation value on the right hand side of (\ref{esa2}) using the cumulant expansion. Recall that the cumulant expansion for an observable $X$ takes the form
\begin{align}
    \log \< \exp(X)\> = \sum_{m=1}^\infty \frac{\kappa_m}{m !}\,,
\end{align}
where $\kappa_m$ is the $m$th cumulant of $X$, e.g.
\begin{align}
\kappa_1 &= \<X\>, \quad \kappa_2 = \<X^2\> - \<X\>^2, \nonumber \\
\kappa_3 &=\<X^3\> - 3\<X^2\>\<X\> + 2\<X\>^3, ...
\end{align}
Now consider the case of interest: $X = - \int_0^\beta d\tau [W(\theta(\tau)) - H_0]$. In this case, it is easy to see that the $m$th cumulant $\kappa_m$ is of order $L (\Delta \tau)^m$ since the integrand vanishes everywhere except in a time interval of order $\Delta \tau$. Hence, to leading order in $\Delta \tau$, we only need to keep the first cumulant $\kappa_1$, i.e. $\log \< \exp(X)\> \approx \<X\>$. In this way, we obtain
\begin{equation}
    \mathcal{S}_n[\theta] \approx -\log\tr[e^{-\beta H_0}] +\left\langle  \int_{0}^{\beta} \D{\tau}[W(\theta(\tau))-H_0]
    \right\rangle_\beta \,,
    \label{eq:snred0}
\end{equation}
up to an error of order $\mathcal{O}(L (\Delta\tau)^2)$. To compute the expectation value, we note that
\begin{align}
    W(\theta)=-\sum_{j=1}^L  &[\sigma^z_j \sigma^z_{j+1}\cos^2 \theta+ \sigma^y_j \sigma^y_{j+1}\sin^2 \theta \nonumber \\
&+(\sigma^z_j\sigma^y_{j+1}+\sigma^y_j\sigma^z_{j+1})\cos \theta\sin\theta]\,.
    \label{wtheta}
\end{align}
Only the first term has a nonzero expectation value, so we derive
\begin{align}
    \<W(\theta) - H_0\>_\beta &= \sin^2 \theta \left< \sum_{j=1}^L \sigma^z_j \sigma^z_{j+1}\right>_\beta \nonumber \\
    &\approx L \tanh \beta \sin^2 \theta\,,
    \label{finite_size_approx}
\end{align}
where the second line follows from the approximation $\left< \sigma^z_j \sigma^z_{j+1}\right>_\beta \approx \tanh \beta$, which holds up to a finite size correction which is exponentially small in $L$. Substituting into (\ref{eq:snred0}) above, we obtain
\begin{equation}\label{theo4}
   \mathcal{S}_n[\theta]=-\log \tr[e^{-\beta H_0}]+L \tanh\beta\int_0^\beta\D{\tau} \sin^2\theta(\tau)\,.
\end{equation}

We can now write down the desired ``reduced'' action: dropping the first term (which is just an overall constant) and adding in the kinetic term in (\ref{actiona}), we obtain
\begin{equation}
   \mathcal{S}_{\rm reduced}[\theta]\equiv L\tanh\beta\int_0^\beta\D{\tau} \sin^2\theta(\tau)+\frac{L^\alpha}{4\lambda}\int_0^\beta \D{\tau} \dot{\theta}(\tau)^2\,.
   \label{reducedactiona}
\end{equation}
Let us now pause to understand the precise meaning of this result. At a qualitative level, we have derived an action $\mathcal{S}_{\rm reduced}[\theta]$ with two properties: (i) $\mathcal{S}_{\rm reduced}[\theta]$ is local in time, and (ii) $\mathcal{S}_{\rm reduced}[\theta]$ approximately agrees with $\mathcal{S}^{\rm tot}_{n}[\theta]$ on all paths obeying our assumptions (\ref{instantonbc}-\ref{instantonstruct}). More quantitatively, we can see that the two approximations in Eqs.~(\ref{approx}) and (\ref{eq:snred0}) both  contribute errors of order $\mathcal{O}(L \Delta \tau^2)$ to the action, while our only other approximation, namely (\ref{finite_size_approx}), contributes an even smaller error which is exponentially small in $L$. Hence $\mathcal{S}_{\rm reduced}[\theta]$ and $\mathcal{S}^{\rm tot}_{n}[\theta]$ must agree with one other up to terms of order $\mathcal{O}(L \Delta \tau^2)$. By comparison, the actions themselves are of order $\mathcal{O}(L \Delta \tau)$, so the difference between the two actions can indeed be neglected in the large $L$ limit assuming $\Delta \tau \rightarrow 0$.

\begin{figure}[t]
\begin{center}
\includegraphics[width=0.5\linewidth]{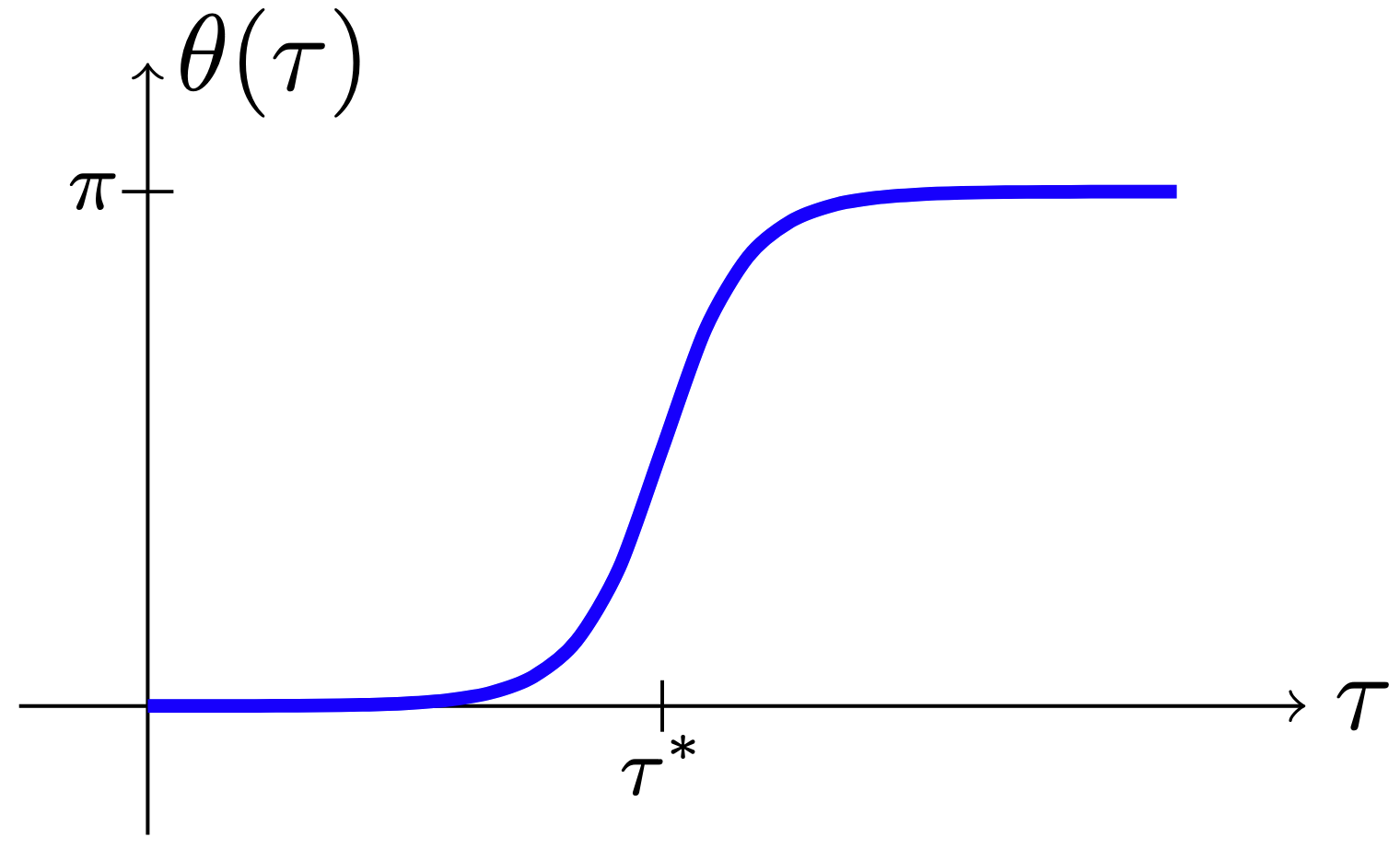}
\caption{Sketch of the minimal path $\bar{\theta}_n(\tau; \tau^*)$ for large but finite $L$.} 
\label{instanton}
\end{center}
\end{figure}

The next step is to minimize $\mathcal{S}_{\rm reduced}[\theta]$, subject to the boundary conditions $\bar{\theta}_n(0)=0$ and $\bar{\theta}_n(\beta)=\pi$. This will give us the minimal path $\bar{\theta}_n$ in the limit $L \rightarrow \infty$.\footnote{Here we are using the fact that $\mathcal{S}_{\rm reduced}$ is \emph{real} so minimizing $\mathcal{S}_{\rm reduced}$ is the same as minimizing $ \Re( \mathcal{S}_{\rm reduced})$.} 
To this end, note that the classical equation of motion for (\ref{reducedactiona}) is
\begin{equation}
     \frac{L^\alpha}{2\lambda}\ddot{\theta}=L\tanh\beta\sin(2\theta)\,,
     \end{equation}
so that
\begin{equation}
    E = \frac{L^\alpha}{4\lambda} \dot{\theta}^2 - L \tanh \beta \sin^2 \theta
\end{equation}
is a constant of motion. In our case, the boundary conditions $\theta(0) = 0$ and $\theta(\beta) = \pi$ correspond to a vanishing $E$ in the $L \rightarrow \infty$ limit, so $\theta$ obeys the first order equation 
\begin{equation}
    \dot{\theta}=2\sqrt{\lambda \tanh\beta}L^{\frac{1-\alpha}{2}}\sin \theta\,.
\label{firstordereq}
\end{equation}
Solving this equation, we obtain a one-parameter family of minimal paths of the form
\begin{equation}
    \bar{\theta}_n(\tau; \tau^*) = 2 \arctan\left(e^{2 \sqrt{\lambda \tanh \beta} L^{\frac{1 -\alpha}{2}} (\tau-\tau^*)} \right),
    \label{thetanform}
\end{equation}
where $\tau^* \in (0, \beta)$ parameterizes the instanton tunneling time. (See Fig.~\ref{instanton}).

This result deserves a few comments. First, notice that these paths do in fact have the step-like structure assumed in Eq.~(\ref{instantonstruct}) with $\Delta \tau \sim \lambda^{-\frac{1}{2}} L^{\frac{\alpha-1}{2}}$. Furthermore, these paths obey the boundary conditions (\ref{instantonbc}) in the $L \rightarrow \infty$ limit. This is an important consistency check on our assumptions (\ref{instantonbc}-\ref{instantonstruct}). 

Another important point is that these paths all have approximately the same action in the limit of large $L$. More precisely, if we let $S(\tau^*)$ denote the action $S(\tau^*) = \mathcal{S}_n^{\rm{tot}}[\bar{\theta}_n(\tau; \tau^*)]$, then for any $\tau_1^*, \tau_2^* \in (0, \beta)$, the difference $S(\tau_1^*) - S(\tau_2^*)$ is exponentially small in $\Delta \tau^{-1} \sim \lambda^{\frac{1}{2}} L^{\frac{1-\alpha}{2}}$. This (nearly exact) degeneracy means that all of these paths contribute equally in the saddle point approximation below.

To complete our derivation, we need to check that the above paths $\bar{\theta}_n$ obey the saddle point condition $\frac{\delta \mathcal{S}^{\rm tot}_{n}}{\delta \theta} = 0$. As we mentioned earlier, this will then imply that these paths dominate the path integral, since $\bar{\theta}_n$ minimizes $\Re( \mathcal{S}^{\rm tot}_{n})$ by construction.

To check the saddle point condition, we note that, to leading order in $L$, the variation 
$\delta S_n$ about the path $\bar{\theta}_n$ is simply:
\begin{align}
    \delta \mathcal{S}_n = \int_0^\beta d\tau \ \< W'(\bar{\theta}_n)\>_\beta \cdot \delta \theta(\tau) + i \<M^x\>_\beta \cdot \delta \theta(\beta)
\end{align}
Here, we are using (\ref{expsaz}) together with the fact that $W(\bar{\theta}_n) \approx H_0$ except for a time interval of length $\Delta \tau \sim \lambda^{-\frac{1}{2}} L^{\frac{\alpha-1}{2}}$. Next, we observe that $\<M^x\>_\beta = 0$. Hence
$\frac{\delta \mathcal{S}_{n}}{\delta \theta} =\<W'(\bar{\theta}_n)\>_\beta$. Including the kinetic term in $\mathcal{S}^{\rm tot}_{n}$ gives $\frac{\delta \mathcal{S}^{\rm tot}_{n}}{\delta \theta} = \frac{\delta \mathcal{S}_{\rm reduced}}{\delta \theta} = 0$, as we wished to show.  

\subsubsection{Evaluating $s_\beta$}
Having found the minimal paths for the numerator and denominator, we are now ready to evaluate $s_\beta$. The first step is to apply the saddle point approximation which gives 
\begin{align}
s_\beta \approx 2\beta K \frac{e^{-\mathcal{S}_{n0}}}{e^{- \mathcal{S}_{d0}}}\,,
\label{sbetasad}
\end{align}
where $\mathcal{S}_{n0}, \mathcal{S}_{d0}$ are the actions of the minimal paths in the numerator and denominator and where $K$ is a factor that comes from evaluating determinants and will be analyzed below. Here, the factor of $2$ comes from the fact that $\mathcal{S}_n$ has two minimal paths, namely $\bar{\theta}_n(\tau)$ and $-\bar{\theta}_n(\tau)$. Likewise, the factor of $\beta$ comes from the fact that  $\bar{\theta}_n$ is actually a one-parameter family of (nearly degenerate) minimal paths, parameterized by $\tau^* \in (0,\beta)$. (More formally, the factor of $\beta$, which is discussed in detail in Ref.~\onlinecite{coleman1988aspects}, originates from the fact that the Hessian of $\mathcal{S}_n$ at the saddle point has an eigenvalue which is very close to zero, and in particular much smaller than the inverse of the large parameter $L$ controlling the saddle point approximation.  
This eigenvalue corresponds to translating the instanton time $\tau^*$. To correctly account for this quasi-zero mode, one needs to integrate over the parameter $\tau^*$, giving the factor of $\beta$.)

Now we proceed to evaluate the minimal actions $\mathcal{S}_{n0}, \mathcal{S}_{d0}$ as well as the determinant factor $K$. We start with $\mathcal{S}_{d0}$. Directly substituting the minimal path $\bar{\theta}_d$ into the action $\mathcal{S}^{\rm tot}_{d}$, we obtain
\begin{align}
\mathcal{S}_{d0} &= \mathcal{S}^{\rm tot}_{d}[\bar{\theta}_d] \nonumber \\
&= - \log\tr[e^{-\beta H_0}]\,.
\label{sd0}
\end{align}
Likewise, to find $\mathcal{S}_{n0}$, we note that
\begin{align}
\mathcal{S}_{n0} &= \mathcal{S}^{\rm tot}_{n}[\bar{\theta}_n] \nonumber \\
&= \mathcal{S}_{\rm reduced}[\bar{\theta}_n] - \log\tr[e^{-\beta H_0}]\,,
\end{align}
where the second term is the constant that we dropped when we wrote down the reduced action (\ref{reducedactiona}). Straightforward calculation gives
\begin{align}
    \mathcal{S}_{\rm reduced}[\bar{\theta}_n]&= \frac{L^{\alpha}}{2\lambda} \int_0^\beta d\tau \dot{\bar{\theta}}_n^2 \nonumber \\
    &= \frac{L^{\alpha}}{2\lambda} \int_0^\pi d\theta \left(2\sqrt{\lambda \tanh\beta}L^{\frac{1-\alpha}{2}} \sin \theta \right) \nonumber \\
    &= 2L^{\frac{1+\alpha}{2}}\sqrt{\frac{\tanh\beta}{\lambda}}\,,
    \label{sn0eval}
\end{align}
where the second equality follows from (\ref{firstordereq}). Hence
\begin{align}
\mathcal{S}_{n0} = 2L^{\frac{1+\alpha}{2}}\sqrt{\frac{\tanh\beta}{\lambda}} - \log\tr[e^{-\beta H_0}]\,.
\label{sn0}
\end{align}
Combining the above expressions for $\mathcal{S}_{d0}$ and $\mathcal{S}_{n0}$, we can now compute the desired ratio $\frac{e^{-\mathcal{S}_{n0}}}{e^{- \mathcal{S}_{d0}}}$:
\begin{align}
\frac{e^{-\mathcal{S}_{n0}}}{e^{- \mathcal{S}_{d0}}} = \exp \left(-2L^{\frac{1+\alpha}{2}}\sqrt{\frac{\tanh\beta}{\lambda}} \right)\,.
\label{ratio0}
\end{align}
Below, it will be important to remember that Eq.~(\ref{ratio0}) is only an approximate formula, correct to leading order in $L$. In particular, recall that when we computed $\mathcal{S}_{\rm reduced}$ and hence $\mathcal{S}_{n0}$, we made several approximations of order $\mathcal{O}(L \Delta\tau^2) = \mathcal{O}(L^\alpha)$; hence the right hand side of (\ref{ratio0}) is only correct up to multiplicative corrections of order $\exp[\mathcal{O}(L^\alpha)]$. 

We now move on to the determinant factor $K$. For this, we use the prescription from Ref.~\onlinecite{coleman1988aspects}, which gives
\begin{equation}
K = c \left| {\frac{\det \mathcal{H}_d}{\det' \mathcal{H}_n}} \right|^{1/2}\,,
\end{equation}
where $c$ is the normalization constant
\begin{equation}
c = \left( \frac{1}{2\pi}\int_0^\beta d\tau \dot{\bar{\theta}}_n^2 \right)^{1/2}\,,
\end{equation}
and where $\mathcal{H}_n$ and $\mathcal{H}_d$ are the Hessians of $\mathcal{S}^{\rm tot}_{n}$ and $\mathcal{S}^{\rm tot}_{d}$, evaluated on the minimal paths. Here the notation $\det'$ means that the determinant is evaluated after removing the (quasi-)zero mode of $\mathcal{H}_n$ associated with shifting $\tau^*$.

In our case, the normalization constant $c$ can be evaluated as in (\ref{sn0eval}), giving
\begin{align}
c = \sqrt{\frac{2}{\pi}} L^{\frac{1-\alpha}{4}} (\lambda \tanh\beta)^{1/4}\,.
\label{c}
\end{align}
As for the determinant ratio, this is not easy to evaluate due to the fact that $\mathcal{S}^{\rm tot}_{n,d}[\theta]$ are non-local functionals of $\theta(\tau)$. Nevertheless, simple arguments suggest that the ratio should scale with $L$ like
\begin{equation}\label{ratiodetermscaling}
  \frac{\det \mathcal{H}_{d}}{\det' \mathcal{H}_{n}} \propto L \,,
\end{equation}
to leading order in $L$ (see Appendix~\ref{hessiannegligible} for a derivation).

Putting this together, we conclude that $K$ is given by
\begin{equation}
K = \sqrt{\frac{2}{\pi}} L^{\frac{3-\alpha}{4}} (\lambda \tanh\beta)^{1/4} \cdot f(\beta, \lambda)
\label{K}\,.
\end{equation}
where $f(\beta, \lambda)$ denotes the proportionality factor in (\ref{ratiodetermscaling}). At this point, it is tempting to simply substitute our expressions for $\frac{e^{-\mathcal{S}_{n0}}}{e^{-\mathcal{S}_{d0}}}$ and $K$ into (\ref{sbetasad}) to obtain a formula for $s_\beta$. However, this is not quite right: recall that the expression for $\frac{e^{-\mathcal{S}_{n0}}}{e^{- \mathcal{S}_{d0}}}$ given in Eq.~(\ref{ratio0}) is only correct up to a multiplicative factor of $\exp[\mathcal{O}(L^\alpha)]$. Comparing this multiplicative factor with our formula for $K$ (\ref{K}), we see that the factor of $K$ is even smaller, so it should be dropped for consistency. Dropping this factor, we derive the desired formula for $s_\beta$:
\begin{align}
\log \left( \frac{s_\beta}{\beta} \right) = -2L^{\frac{1+\alpha}{2}}\sqrt{\frac{\tanh\beta}{\lambda}}\,.
\label{sbeta}
\end{align}

To understand the significance of (\ref{sbeta}), we now tabulate the various approximations that we made in deriving it. First, in our calculation of $\mathcal{S}_{n 0}$, there were several steps where we neglected terms of order $\mathcal{O}(L^\alpha)$, as mentioned earlier. Second, when we used the saddle point approximation in (\ref{sbetasad}), we neglected paths with more than one instanton. These paths will give corrections to (\ref{sbeta}) that are of order $[\beta \exp(-\mathcal{O}( L^{\frac{1+\alpha}{2}}))]^n$ where $n \geq 2$ is the number of instantons (since each additional instanton contributes a phase space of order $\beta$ and an action of order $L^{\frac{1+\alpha}{2}}$). 

What is most important for our purposes is that, in the limit $L \rightarrow \infty$, all of these errors are \emph{subleading} compared with the single instanton term in (\ref{sbeta}) (assuming $\alpha < 1$). Therefore (\ref{sbeta}) gives the exact asymptotic scaling of $\log(s_\beta)$ as $L \rightarrow \infty$.

Let us finally mention the importance of the regime $\alpha<1$ for our saddle point calculation: the key point is that the instanton solution we find is of size $\Delta\tau=\mathcal{O}(L^{\frac{\alpha-1}{2}})$. This can be consistent with the assumptions mentioned at the beginning of Section \ref{minimalpath} only in the case $\alpha<1$.

\subsection{Computing the ground state splitting $\delta$}

Having found $s_\beta$ (\ref{sbeta}), we are now ready to compute the main quantity of interest, namely the  ground state splitting $\delta$. We begin by computing the \emph{free} energy splitting $\delta F_\beta$ using (\ref{deltasplit}). Expanding (\ref{deltasplit}) to lowest order in $s_\beta$ gives
\begin{equation}
\delta F_\beta = \frac{2 s_\beta}{\beta}\,.
\end{equation}
Taking the logarithm of both sides and substituting our formula for $s_\beta$ (\ref{sbeta}), we obtain
\begin{equation}
\log \delta F_\beta = -2L^{\frac{1+\alpha}{2}}\sqrt{\frac{\tanh\beta}{\lambda}}\,.
\end{equation}
Finally, taking the limit $\beta \rightarrow \infty$, we derive a formula for the ground state splitting $\delta$:
\begin{equation}
\log \delta = -\frac{2}{\sqrt{\lambda}} L^{\frac{1+\alpha}{2}}\,.
\end{equation}
Like our formula for $s_\beta$ (\ref{sbeta}) this result is asymptotically exact in the sense that the ratio of the left and right sides approaches $1$ in the limit $L \rightarrow \infty$.

\section{Quantum rotor model with power-law interactions \label{sec:rotor}}

In the previous section, we showed that the ground state splitting depends on $L$ in a stretched exponential fashion for the Hamiltonian (\ref{H}) with all-to-all interactions, $f(r) = \frac{1}{4L^\alpha}$. We now investigate whether similar behavior occurs for power-law interactions, $f(r)\sim 1/r^\alpha$.

Unfortunately, it is not obvious how to generalize our calculation to the power-law case. To see the problem, suppose we decouple the $f(|i-j|) \sigma^x_i \sigma^x_j$ term using a Hubbard-Stratonovich transformation as we did in Eq.~\ref{eq:hubbstrat}. To do this, we need to introduce $L$ fields $\varphi_1, ..., \varphi_L$, each of which couple to a different spin $\sigma_1^x,...,\sigma_L^x$. The resulting expression for $s_\beta$ then involves a path integral over $L$ degrees of freedom, rather than just a single degree of freedom. Furthermore, there is no large parameter in the action, like the coefficient of $L^\alpha$ that appears in the kinetic term in (\ref{actiona}) or the factor of $L$ that is implicit in the $\mathcal{S}_{n,d}[\theta]$ terms (due to their extensive nature). As a result, the saddle-point approximation is not justified and it is not obvious how to proceed. 

To address this issue, we now introduce a model that is analogous to the Ising spin chain \eqref{H}, but that allows for a semi-classical limit at fixed $L$. This will allow us to perform a saddle-point approximation in a well-controlled setting; in fact, we will be able to apply the standard instanton method introduced by Coleman~\cite{coleman1988aspects}.

\subsection{Hamiltonian}
Our model consists of a periodic chain of $L$ quantum rotors. Each rotor is described by a number operator $n$ and a phase operator $\theta$ obeying the canonical commutation relation, $[\theta, n] = i$. Here, the phase $\theta$ is an angular variable, defined modulo $2 \pi$, while the canonically conjugate number operator $n$ has integer eigenvalues. The corresponding number eigenstates $\{|p\>, p \in \mathbb{Z}\}$ form a complete orthonormal basis for the Hilbert space for each rotor. In this basis, the number and phase operators act as $n|p\> = p |p\>$ and $e^{\pm i \theta} |p\> = |p \pm 1\>$. Alternatively, one can work in the phase eigenstate basis, $\{|\theta\>, \theta \in [0, 2\pi]\}$ defined by $\<p | \theta\> = e^{-i p \theta}$. 

Denoting the number and phase operators of the $j$th rotor by $n_j$ and $\theta_j$, our Hamiltonian takes the form
\begin{align}
    H &= \frac{1}{g} \sum_{j=1}^L [1 -\cos(\theta_j-\theta_{j+1})] + \frac{1}{g}\sum_{j=1}^L U(\theta_j) \nonumber \\
    &+\frac{g}{2}\sum_{j=1}^L n_j^2 + g\lambda \sum_{i, j=1}^L  f(|i-j|) n_i n_j\,.
    \label{confinedh}
\end{align}
where $U(\theta)$ is the periodic potential defined by
\begin{equation}
\label{u}
U(\theta) = \frac{1}{2}(|\theta| - \pi/2)^2, \quad -\pi \leq \theta \leq \pi\,.
\end{equation}
This Hamiltonian is similar to standard Josephson-junction-type models~\cite{Fisher-Grinstein88} except for the $U(\theta)$ term, which breaks the $U(1)$ number conservation symmetry generated by $\sum_j n_j$ down to the $\mathbb{Z}_2$ symmetry $S$ (\ref{Srotor}) defined below. (Here, the meaning of $\cos(\theta_j-\theta_{j+1})$ and $U(\theta_j)$ as operators is obtained by writing the Fourier decomposition of these functions in $e^{i\theta_{j}}$ and $e^{i \theta_{j+1}}$, and replacing these terms by the ladder operators defined above by $e^{\pm i\theta}|p\>=|p\pm 1\>$). 

Let us explain the various parameters in this model. The parameter $g > 0$ controls our semi-classical expansion: we will study the model in the limit $g \rightarrow 0$ with $g$ playing the role of $\hbar$. The parameter $\lambda \geq 0$ has the same meaning as before: it tunes the strength of the long-range interaction. As for $f(r)$, this function describes the form of the long-range interaction. We will eventually specialize to the case where $f(r)$ is the power-law interaction given in (\ref{periodpower}), but for now, we will allow $f(r)$ to be arbitrary, with the only restrictions being that $|f(r)| \leq 1$ and $f(r) = f(L-r)$ (which is needed for consistency with periodic boundary conditions).

As we mentioned earlier, we designed the model (\ref{confinedh}) so that it is a close analog of the Ising spin chain (\ref{H}). To see this analogy, notice that the Hamiltonian (\ref{confinedh}) is invariant under the $\mathbb{Z}_2$ symmetry defined by
\begin{equation}
    S= e^{i \pi \sum_{j=1}^L n_j}\,,
    \label{Srotor}
\end{equation}
which implements a uniform shift $\theta_j \rightarrow \theta_j +\pi$. Furthermore, notice that the $n_j$'s that appear in the long-range interaction term
$\sum_{i,j} n_i n_j f(|i-j|)$ in (\ref{confinedh}) are generators of the symmetry $S$, just like the $\sigma^x_j$'s that appear in the corresponding term $ \sum_{i,j} \sigma_i^x \sigma_j^x f(|i-j|)$ in the Ising spin chain (\ref{V}). [Here, we are using the fact that the Ising symmetry (\ref{Sising}) can be written as $S = e^{i \pi/2 \sum_j \sigma_j^x}$ similarly to (\ref{Srotor}).] Finally, notice that
the rotor model (\ref{confinedh}) spontaneously breaks the above $\mathbb{Z}_2$ symmetry (\ref{Srotor}) in the $g \rightarrow 0$ limit. Indeed, the potential $U(\theta)$ has two minima at $\theta = \pm \pi/2$, while the cosine interaction term $\cos(\theta_j - \theta_{j+1})$ locks neighboring rotors in the same minimum of $U(\theta)$. Thus, when $\lambda = 0$ and $g \rightarrow 0$, the rotor model (\ref{confinedh}) has two degenerate classical ground states at $\theta_j = \pm \pi/2$ -- just like the two degenerate ground states of (\ref{H0}). Putting this all together, we can view the rotor model (\ref{confinedh}) as a semi-classical analog of the Ising spin chain (\ref{H}).

\subsection{Instanton method}
We now compute the ground state splitting of the rotor model (\ref{confinedh}) in the semi-classical limit $g \rightarrow 0$. We accomplish this using the instanton method introduced by Coleman~\cite{coleman1988aspects}. Below we outline the main steps in the calculation. 

The first step is to write down the imaginary time path integral corresponding to the Hamiltonian (\ref{confinedh}). Let $\bm{\theta}^{\rm ini}$ and $\bm{\theta}^{\rm fin}$ be a set of initial and final values of the phases $\theta_1,...,\theta_L$. As in the standard path integral representation of non-relativistic quantum mechanics, the matrix elements of $e^{-\beta H}$ in the $|\theta_1,...,\theta_L\>$ basis are given by the following path integral:
\begin{equation}
    \langle \bm{\theta}^{\rm ini}|e^{-\beta H}|\bm{\theta}^{\rm fin}\rangle=\int \mathcal{D}[\bm{\theta}]e^{-\mathcal{S}[\bm{\theta}]/g}\,,
\end{equation}
where the $L$-dimensional trajectories $\bm{\theta}(\tau)$ go from $\bm{\theta}^{\rm ini}$ at time $0$ to $\bm{\theta}^{\rm fin}$ at time $\beta$, and where the (imaginary time) action takes the form
\begin{align}
    \mathcal{S}[\bm{\theta}]=\int_{0}^{\beta}\D{\tau}& \biggl[ \frac{1}{2}\sum_{i,j=1}^L \dot{\theta}_i(\tau) M_{ij}\dot{\theta}_j(\tau)+\sum_{j=1}^L U(\theta_j(\tau)) \nonumber \\
    &+ \sum_{j=1}^L [1-\cos(\theta_j(\tau)-\theta_{j+1}(\tau))] \biggr]\,,
    \label{actionsemi}
\end{align}
with the mass matrix $M_{ij}$ given by 
\begin{equation}
    M^{-1}_{ij}=\delta_{ij}+2\lambda f(|i-j|)\,.
\end{equation}

In what follows, we will assume that $M_{ij}^{-1}$ is \emph{positive-definite} so that the Hamiltonian (\ref{confinedh}) and the action (\ref{actionsemi}) are bounded from below. In general, this positive-definiteness is guaranteed to hold for sufficiently small $\lambda$. More specifically, when $f(r)$ is the power-law interaction (\ref{periodpower}), one can check that $M_{ij}^{-1}$ is positive-definite for $0 \leq \lambda \leq \lambda_0$ where $\lambda_0$ is of order $1$.

According to Ref.~\onlinecite{coleman1988aspects},  the ground state splitting $\delta$ is controlled by the ``instanton'' path $\bar{\bm{\theta}}(\tau)$. This path is defined by two properties: (i) $\bar{\bm{\theta}}$ obeys the boundary conditions
\begin{equation}
\bar{\theta}_i(\tau = -\infty) = -\frac{\pi}{2}, \quad \quad \bar{\theta}_i(\tau = \infty) = \frac{\pi}{2},
\label{instantonbc2}
\end{equation}
and (ii) $\bar{\bm{\theta}}$ minimizes the imaginary time action (\ref{actionsemi}). 
The physical interpretation of the instanton is that it describes a tunneling process connecting the two classical ground states at $\theta_i = -\frac{\pi}{2}$ and $\theta_i = \frac{\pi}{2}$. 
Once we have the instanton path, we can compute the leading behavior of the splitting $\delta$, in the limit $g \rightarrow 0$ using the formula 
\footnote{There is an extra factor of $2$ compared to Ref.~\onlinecite{coleman1988aspects} because $\theta$ is defined modulo $2\pi$ and therefore there are two instanton solutions going from $-\pi/2$ to $\pi/2$, namely $\bar{\theta}_i(\tau)$ and $\pi - \bar{\theta}_i(\tau)$.}
\begin{equation}
    \delta=4 K e^{-\mathcal{S}[\bar{\bm{\theta}}]/g}\,,
    \label{deltasad}
\end{equation}
where 
\begin{equation}
K =  c\left| \frac{\det \mathcal{H}^{(1)}}{\det' \mathcal{H}^{(2)}} \right|^{1/2}\,
\label{Kformrot}
\end{equation}
is a factor coming from fluctuation determinants, and $c$ is the normalization constant
\begin{equation}
c = \left( \frac{1}{2\pi g}\int_0^\beta d\tau \sum_{i=1}^L \dot{\bar{\theta}}_i^2 \right)^{1/2}\,.
\end{equation}
Here $\mathcal{H}^{(1)}$ and $\mathcal{H}^{(2)}$ are Hessians of the action (\ref{actionsemi}) evaluated along two paths, namely (1) the trivial path $\bm{\theta}(\tau)=-\frac{\pi}{2}$ and (2) the instanton path $\bm{\theta}=\bar{\bm{\theta}}$.
As in the previous section, the notation $\det'$ means that the determinant is evaluated after removing the zero mode associated with translating the instanton in time.

Where does the above formula (\ref{deltasad}) come from? A detailed derivation is given in Ref.~\onlinecite{coleman1988aspects} but the basic idea is to consider the matrix element $\<-\pi/2| e^{-\beta H} | \pi/2\>$ in the limit $g \rightarrow 0$ and $\beta \rightarrow \infty$. Using the saddle point approximation, one can see that this matrix element is controlled by the instanton path $\bar{\bm{\theta}}$ and small fluctuations about this path. At the same time, we can relate the matrix element $\<-\pi/2| e^{-\beta H} | \pi/2\>$, or more precisely, its ratio with $\<-\pi/2| e^{-\beta H} |-\pi/2\>$, to the ground state splitting $\delta$ using a spectral decomposition of $e^{-\beta H}$. Combining these two calculations gives the above formula (\ref{deltasad}). 

\subsection{Computation of ground state splitting}
We now use the instanton method outlined in the previous section to compute the ground state splitting $\delta$ in the semi-classical limit $g \rightarrow 0$. The first step is to find the instanton path $\bar{\bm{\theta}}$. The classical equation of motion for the action (\ref{actionsemi}) is 
\begin{equation}    
\sum_j M_{ij}\ddot{\theta}_j=U'(\theta_i)+  \sin(\theta_i-\theta_{i+1})+  \sin(\theta_i-\theta_{i-1})\,.
\label{eomrotor}
\end{equation}
Given that both the action (\ref{actionsemi}) and the boundary conditions (\ref{instantonbc2}) are invariant under translations, $\theta_i \rightarrow \theta_{i+1}$, it seems reasonable to assume that the instanton solution is also translationally symmetric, i.e. 
\begin{align}
\theta_1 = \theta_2 = ... = \theta_L = \theta
\end{align}
for some some $\theta$. Substituting this ansatz into (\ref{eomrotor}) gives the following (single particle) equation of motion:
\begin{equation}\label{derivative}
    \ddot{\theta}=\frac{U'(\theta)}{m_0}\,,
\end{equation}
where $m_0$ is the effective mass
\begin{equation}
    m_0=\frac{1}{1+2\lambda\sum_{r=0}^{L-1} f(r)}\,.
\end{equation}
Next, we note that
\begin{align}
E = \frac{m_0}{2} \dot{\theta}^2 - U(\theta)
\end{align}
is a constant of motion for this equation. Our boundary conditions $\theta(\tau = -\infty) = -\pi/2$ and $\theta(\tau = \infty) = \pi/2$ imply that $E = 0$ so $\theta$ obeys the first order equation
\begin{equation}
    \dot{\theta}=\sqrt{\frac{2U(\theta)}{m_0}}\,,
\end{equation}
Solving this equation with $U$ defined as in (\ref{u}) gives a one-parameter family of instanton paths of the form
\begin{equation}
\bar{\theta}(\tau; \tau^*) = \frac{\pi}{2} \text{sgn}(\tau - \tau^*)  \left[1 - e^{-\frac{1}{\sqrt{m_0}} |\tau - \tau^*|} \right]
\label{instantonrot}
\end{equation}
where $\tau^*$ parameterizes the center of the instanton.

Having found the instanton path $\bar{\bm{\theta}}$, our next task is to evaluate the corresponding instanton action $\mathcal{S}[\bar{\bm{\theta}}]$. Direct calculation gives:
\begin{align}
    \mathcal{S}[\bar{\bm{\theta}}]&=L m_0 \int_{-\infty}^{\infty} \D{\tau} \dot{\theta}^2 \nonumber \\
    &= L\int_{-\pi/2}^{\pi/2} \D{\theta} \sqrt{2m_0 U(\theta)} \nonumber \\
 &=L \frac{\pi^2}{4} \sqrt{m_0}\,.
\label{Sinstantonrot}
\end{align}
Likewise, we find 
\begin{align}
c &= \left(\frac{L}{2\pi g} \int_{-\infty}^{\infty} \D{\tau} \dot{\theta}^2 \right)^{1/2} \nonumber \\
&= \left( \frac{L \pi}{8 g \sqrt{m_0}}\right)^{1/2}
\end{align}

Lastly we need to compute the ratio of fluctuation determinants $\frac{\det \mathcal{H}^{(1)}}{\det' \mathcal{H}^{(2)}}$ that appears in $K$ (\ref{Kformrot}).
We carry out this calculation in Appendix~\ref{apphess}, and we find
\begin{align}
\left| \frac{\det \mathcal{H}^{(1)}}{\det' \mathcal{H}^{(2)}} \right| =
2 \prod_{k=1}^{L-1} \left(1-\frac{1}{\sqrt{1+4 \sin^2(\tfrac{\pi k}{L})}}\sqrt{\frac{m_0}{m_k}} \right)^{-1}\,.
\label{detratiorot}
\end{align}
where the $m_k$'s are defined as
\begin{equation}\label{massmodes1}
    \frac{1}{m_k}=1+2\lambda \sum_{r=0}^{L-1} \cos(2\pi k r/L)f(r)\,.
\end{equation}
Substituting (\ref{Sinstantonrot}) and (\ref{detratiorot}) into the formula for the splitting (\ref{deltasad}) and taking the log of both sides gives
\begin{align}
    \log \delta &=-\frac{L \pi^2 \sqrt{m_0}}{4 g} + \frac{1}{2} \log \left( \frac{ 4L \pi }{ g \sqrt{m_0}} \right) \nonumber \\
    &-\frac{1}{2}\sum_{k=1}^{L-1}\log\left(1-\frac{1}{\sqrt{1+4 \sin^2(\tfrac{\pi k}{L})}}\sqrt{\frac{m_0}{m_k}}  \right),
    \label{exphbar}
\end{align}
This expression is the desired semi-classical formula for the splitting $\delta$. The corrections to this formula are of order $o(g^0)$: i.e. the difference between the left and right hand sides vanishes in the limit $g \rightarrow 0$.

\subsection{Scaling of the splitting $\delta$ with $L$}
We now analyze how $\delta$ scales with $L$ in the limit $L \rightarrow \infty$, assuming that the interaction $f(r)$ is a power-law. More precisely, we assume that $f(r)$ is a power-law with respect to an appropriate distance measure for our periodic geometry: 
\begin{align}
    f(r) = \begin{cases} \frac{c_\alpha}{\min(r, L-r)^\alpha}, & r \neq 0 \\
    0, & r = 0\,,
    \end{cases}
    \label{periodpower}
    \end{align}
where $0 < \alpha < 1$ and $c_\alpha = \frac{1-\alpha}{2^{\alpha}}$. Here, the reason we choose the above value for $c_\alpha$ is because it simplifies some of the formulas below. In particular, for this choice of $c_\alpha$, we have
\begin{align}
\sum_{r=0}^{L-1} f(r) = L^{1-\alpha}+\mathcal{O}(L^{-\alpha})\,,
\end{align}
in the large $L$ limit. Hence the mass $m_0$ behaves as
\begin{equation}
m_0 = \frac{L^{\alpha-1}}{2\lambda}+\mathcal{O}(L^{\alpha-2})
\label{m0Ninfty}
\end{equation}
as $L \rightarrow \infty$.

Fixing $f(r)$ as in (\ref{periodpower}), our task is to understand how $\delta$ scales with $L$ in the limit $L \rightarrow \infty$. We will follow a naive approach: we will simply analyze the $L \rightarrow \infty$ limit of our semi-classical formula (\ref{exphbar}). This approach implicitly assumes that the next terms in the semi-classical expansion do not grow any faster with $L$ than the terms in (\ref{exphbar}).

To proceed, we substitute the formula for $m_0$ (\ref{m0Ninfty}) into (\ref{exphbar}) giving
\begin{widetext}
\begin{equation}
   \log \delta  =    - L^{\frac{1+\alpha}{2}} \frac{ \pi^2}{4g\sqrt{2\lambda}}+ \frac{1}{2} \log \left( \frac{4L^{\frac{3-\alpha}{2}}  \pi \sqrt{2\lambda} }{g} \right) 
    -\frac{1}{2}\sum_{k=1}^{L-1}\log\left(1-\frac{1}{\sqrt{2\lambda}}\sqrt{\frac{1 + 2\lambda\sum_{r=0}^{L-1} \cos(2\pi kr/L)f(r)}{1+4 \sin^2(\tfrac{\pi k}{L})}}L^{\frac{\alpha-1}{2}}  \right)\,.
\end{equation}
Next, we note we can make the following approximation in the limit of large $L$, assuming $k$ is of order $L$: 
\begin{align}
\sum_{r=0}^{L-1} \cos(2\pi kr/L)f(r) \approx 2 c_\alpha \sum_{r=1}^\infty \frac{\cos(2\pi kr/L)}{r^\alpha}
\end{align}
Using this approximation and converting the sum over $k$ into an integral gives
\begin{equation}\label{logf}
    \log \delta =-L^{\frac{1+\alpha}{2}}\frac{ \pi^2}{4g\sqrt{2\lambda}}+ \frac{1}{2} \log \left( \frac{4L^{\frac{3-\alpha}{2}} \pi \sqrt{2\lambda} }{g}  \right) 
    -\frac{L}{2}\int_{0}^{1}\D{x}\log\left(1-\frac{1}{\sqrt{2\lambda}} \sqrt{\frac{1+4\lambda c_\alpha\sum_{r=1}^{\infty} \cos(2\pi r x)/r^\alpha}{1+4 \sin^2(\pi x)}}L^{\frac{\alpha-1}{2}}\right).
\end{equation}
Finally, expanding the $\log$ at large $L$, we derive
\begin{equation}\label{logfj}
    \log \delta =-L^{\frac{1+\alpha}{2}}\frac{ \pi^2}{4g\sqrt{2\lambda}}+ \frac{1}{2} \log \left(\frac{4L^{\frac{3-\alpha}{2}} \pi \sqrt{2\lambda}}{g} \right) 
    +\frac{L^{\frac{1+\alpha}{2}}}{2 \sqrt{2 \lambda}}\int_{0}^{1}\D{x}\sqrt{\frac{1+4\lambda c_\alpha\sum_{r=1}^{\infty} \cos(2\pi r x)/r^\alpha}{1+4 \sin^2(\pi x)}}.
\end{equation}
\end{widetext}
Notice that every term in this semi-classical expansion has a coefficient that is either of order $L^{\frac{1+\alpha}{2}}$ or smaller as $L \rightarrow \infty$. Assuming that the corrections to this formula (which necessarily vanish in the limit $g \rightarrow 0$) are also of order $L^{\frac{1+\alpha}{2}}$ or smaller, we conclude that
\begin{align}
\log \delta  = -C(g) L^{\frac{1+\alpha}{2}}\,,    
\label{stretched}
\end{align}
where $C(g)$ is a function of $g$ whose leading behavior in the limit $g \rightarrow 0$ is given by  $C(g) = \frac{\pi^2}{4g\sqrt{2\lambda}}$.
Evidently, the ground state splitting of our rotor model with power-law interactions scales like the same stretched exponential (\ref{deltaformalltoall}) as the Ising spin chain with all-to-all interactions. 

\section{Toy model with all-to-all interactions \label{sec:toy}}
So far we have analyzed the ground state splitting for two variants of the Ising spin chain with power-law interactions (\ref{H}). In this section we analyze a third variant. The advantage of this model is that its ground state splitting can be computed relatively easily, without any path integral techniques. Furthermore this computation can be performed for multiple types of long-range interactions. The main disadvantage of this model is that the Hamiltonian contains a \emph{non-local} term (in addition to long-range spin-spin interactions). As a result, this model might display certain characteristics that are not present for physical topological qubits with long-range interactions.

\subsection{Hamiltonian}

Our toy model is built out of $L$ spin-$1/2$ degrees of freedom, arranged in a 1D chain with periodic boundary conditions.\footnote{Here, we focus on the 1D geometry for simplicity but the model can be defined equally well on any lattice.} The Hamiltonian is
\begin{equation}\label{toymodel}
    H=-L|\psi_+\>\<\psi_+| -L|\psi_-\>\<\psi_-|
    +\frac{\lambda}{L^\alpha}\left( \sum_{j=1}^L \mathcal{O}_j\right)^2\,,
\end{equation}
where  
\begin{equation}
    |\psi_{\pm}\rangle=\frac{1}{\sqrt{2}}\left(|\uparrow \cdots \uparrow\> \pm |\downarrow \cdots \downarrow\> \right)\,,
\end{equation}
are the ferromagnetic ``cat'' states and where $\lambda \geq 0$ and $0 < \alpha < 1$. The operators $\mathcal{O}_1,..., \mathcal{O}_L$ can be chosen arbitarily as long as they obey the following conditions: (i) $\mathcal{O}_j$ is a local Hermitian operator supported near site $j$; (ii) the operator norm $\|\mathcal{O}_j\| \leq 1$; (iii) $\mathcal{O}_j$ commutes with the Ising symmetry $S = \prod_j \sigma^x_j$; and (iv) $\mathcal{O}_j$ satisfies 
\begin{align}
\<\psi_+| \mathcal{O}_j|\psi_+\>= \<\psi_-| \mathcal{O}_j|\psi_-\>=0.
\end{align}
For example, one possible choice for $\mathcal{O}_j$ is $\mathcal{O}_j = \sigma^x_j$. We will study this example (and others) below. 

We note that the above toy model (\ref{toymodel}) is similar in many ways to the Ising spin chain with all-to-all interactions (\ref{ising}) from Sec.~\ref{sec:all-to-all}: like (\ref{ising}), the toy model is Ising symmetric and has all-to-all interactions with a power-law scaling proportional to $\frac{\lambda}{L^\alpha}$. Also, when $\lambda = 0$, the toy model has two exactly degenerate ground states $|\uparrow \cdots \uparrow\>$, $|\downarrow \cdots \downarrow\>$ that spontaneously break the Ising symmetry, just like the Ising spin chain (\ref{ising}). The main difference between the two models is that the symmetry breaking in (\ref{ising}) is induced by local interactions $\sum_{j=1}^L \sigma^z_{j} \sigma^z_{j+1}$ while the symmetry breaking in (\ref{toymodel}) comes from the non-local terms $-L|\psi_+\>\<\psi_+| -L|\psi_-\>\<\psi_-|$. This lack of locality makes the toy model less physical, but it also simplifies the solution of the model and enables us to consider more general all-to-all interactions.

\subsection{Computing the splitting}

We now compute the splitting $\delta$ between the ground states of the toy model (\ref{toymodel}) in the $S = \pm 1$ sectors. Our strategy is to use the special structure of the Hamiltonian $H$. In particular, notice that within each symmetry sector $S = \pm 1$, the Hamiltonian can be written as
\begin{align}
    H = A - L |\psi_\pm\> \<\psi_\pm|\,,
\end{align}
where $|\psi_\pm\> \<\psi_\pm|$ is a rank one projector and $A =\frac{\lambda}{L^\alpha}\left( \sum_{j=1}^L \mathcal{O}_j\right)^2$. Because $H$ has this structure, we can use a linear algebra identity (the so-called ``matrix determinant lemma'') to write the characteristic polynomial $\det_\pm(H-z \mathbbm{1})$ within each $S = \pm 1$ sector as
\begin{align}
    \text{det}_\pm&(H-z \mathbbm{1})= \nonumber \\&\text{det}_\pm(A-z \mathbbm{1}) \cdot \left[1 - L \<\psi_\pm | (A - z \mathbbm{1})^{-1} |\psi_\pm\> \right] \,.
\label{matdetlemma}
\end{align}

We are interested in finding the smallest root of the above characteristic polynomial: this will give us the ground state energy $E_\pm$ within the $S = \pm 1$ sector. To this end, notice that $\det_\pm(A - z \mathbbm{1}) > 0$ for all negative $z$ since $A$ is positive semi-definite. At the same time, we know that $E_\pm$ is negative for large $L$ from variational arguments, since $\langle \psi_{\pm}|H|\psi_{\pm}\rangle=-L+\mathcal{O}(L^{1-\alpha})$. We conclude that $E_\pm$ must be the smallest root of the \emph{second} factor in (\ref{matdetlemma}), i.e.~$E_\pm$ is the smallest $z$ that satisfies
\begin{align}
    1 - L \<\psi_\pm | (A - z \mathbbm{1})^{-1} |\psi_\pm\> = 0\,.
\label{zeq}
\end{align}

In fact, we can sharpen this statement slightly: $E_\pm$ is the \emph{unique} solution to (\ref{zeq}) with $z < 0$. To see this, notice that the left hand side of (\ref{zeq}) is a decreasing function of $z$ for negative $z$, and therefore there is at most one solution for $z < 0$. 

Equivalently, if we introduce a rescaled variable $\eta$, defined by $z = -L\eta $, then $E_\pm$ is given (up to factor of $-L$) by the unique positive zero of
\begin{align}
    f_\pm(\eta) = 1 - L \<\psi_\pm | (A + L\eta \mathbbm{1})^{-1} |\psi_\pm\>\,.
    \label{fpmdef}
\end{align}

To proceed further, we rewrite $f_\pm(\eta)$ as
\begin{align}
     f_\pm(\eta) = 1 - \frac{L^{1+\alpha}}{\lambda} \left \langle \psi_\pm \Bigg | \left(P^2 + \frac{L^{1+\alpha} \eta}{\lambda} \mathbbm{1} \right)^{-1} \Bigg |\psi_\pm \right \rangle \,,
     \label{fpm2}
\end{align}
where
\begin{equation}
    P=\sum_{j=1}^L \mathcal{O}_j\,.
\end{equation}
Next, we use the following operator identity, which holds for any Hermitian operator $P$:
\begin{equation}
    (P^2 + \omega^2 \mathbbm{1})^{-1} = \frac{1}{\omega} \int_{0}^\infty \D{t}e^{-\omega t} \cos(P t)\,.
    \label{operidtoy}
\end{equation}
Substituting (\ref{operidtoy}) into (\ref{fpm2}) with $\omega= \sqrt{\frac{L^{1+\alpha} \eta}{\lambda}}$ gives
\begin{align}
    f_\pm(\eta) = 1 - \sqrt{\frac{L^{1+\alpha}}{\lambda \eta}} \int_{0}^\infty \D{t}e^{-\sqrt{\frac{L^{1+\alpha} \eta}{\lambda}} t} \<\psi_\pm| \cos(P t) |\psi_\pm\>\,.
    \label{fpmform}
\end{align}

At this point, it is useful to consider symmetric and antisymmetric combinations of $f_+, f_-$:
\begin{align}
    f(\eta) \equiv \frac{1}{2}(f_+(\eta) + f_-(\eta)), \qquad g(\eta) \equiv \frac{1}{2}(f_+(\eta) - f_-(\eta))\,.
\end{align}
From (\ref{fpmform}), we have
\begin{align}
    f(\eta) &=  1 - \sqrt{\frac{L^{1+\alpha}}{\lambda \eta}} \int_{0}^\infty \D{t}e^{-\sqrt{\frac{L^{1+\alpha} \eta}{\lambda}} t} \ \<\Uparrow| \cos(P t) |\Uparrow\> \label{fform}\\
    g(\eta) &=  - \sqrt{\frac{L^{1+\alpha}}{\lambda \eta}} \int_{0}^\infty \D{t}e^{-\sqrt{\frac{L^{1+\alpha} \eta}{\lambda}} t} \ \<\Downarrow| \cos(P t) |\Uparrow\>\,, 
        \label{gform}
\end{align}
where we are using the abbreviation $|\Uparrow\rangle \equiv |\uparrow \cdots \uparrow\>$ and $|\Downarrow\rangle \equiv |\downarrow \cdots \downarrow\>$. 

Let us try to understand the limiting behavior of $f, g$ as $L \rightarrow \infty$. We start with $f$. To understand this function, consider the expectation value $\<\Uparrow| \cos(P t) |\Uparrow\>$. Using the fact that $1-x^2/2 \leq \cos(x) \leq 1$, we have the bounds
\begin{align}
    1- \mathcal{O}(L) t^2 \leq \<\Uparrow| \cos(P t) |\Uparrow\>\leq 1\,.
    \label{alphaineq}
\end{align}
Substituting the inequality (\ref{alphaineq}) into (\ref{fform}) and integrating gives
\begin{align}
 1 - \frac{1}{\eta} \leq f(\eta) \leq 1-\frac{1}{\eta} + \mathcal{O}(L^{-\alpha}) \frac{\lambda}{\eta^2}\,.
 \label{fbound}
\end{align}
We conclude that 
\begin{align}
\lim_{L \rightarrow \infty} f(\eta) = 1 - 1/\eta   \,.
\label{limf}
\end{align}
Similarly, we have
\begin{align}
\lim_{L \rightarrow \infty} g(\eta) = 0\,.
\label{limg}
\end{align}
To see this, consider the quantity $\<\Downarrow| \cos(P t) |\Uparrow\>$. We have the bound
\begin{align}
|\<\Downarrow| \cos(P t) |\Uparrow\>|^2 \leq 1 - |\<\Uparrow| \cos(P t) |\Uparrow\>|^2\,,
\label{cosptineq}
\end{align}
which follows from the fact that $\|\cos(P t) |\Uparrow\>\| \leq 1$. Combining (\ref{cosptineq}) and (\ref{alphaineq}) gives the bound
\begin{align}
|\<\Downarrow| \cos(P t) |\Uparrow\>| \leq \mathcal{O}(L^{1/2}) t\,.
\end{align}
Substituting this inequality into (\ref{gform}) and integrating gives
\begin{align}
|g(\eta)| \leq \mathcal{O}(L^{-\alpha/2}) \frac{\sqrt{\lambda}}{\eta^{3/2}}\,,
\end{align}
implying (\ref{limg}).

We are now ready to analyze the large $L$ behavior of the ground state splitting $\delta$. First, notice that the problem of finding the unique positive zero of $f_\pm(\eta)$ is equivalent to finding the unique positive solution to
\begin{align}
    f(\eta) = \mp g(\eta)
    \label{feqg}\,.
\end{align}
Next, notice that $f$ is an increasing function of $\eta$ for $\eta > 0$:  this follows from the fact that $f_\pm$ are increasing for $\eta > 0$, as one can verify by differentiating (\ref{fpmdef}). It follows that $f$ has at most one positive zero. At the same time, we know $f$ has at \emph{least} one positive zero when $L$ is large since $f(\eta) \approx 1 - 1/\eta$. Hence, $f$ has a unique positive zero when $L$ is large, which we will denote by $\eta^*$.

Let $\eta_\pm$ denote the unique positive solution to (\ref{feqg}). Since $g(\eta) \approx 0$ for large $L$, we know that $\eta_\pm \approx \eta^*$. More quantitatively, we can derive the leading behavior of $\eta_\pm - \eta^*$ by solving (\ref{feqg}) to linear order in $g$:
\begin{align}
    \eta_\pm = \eta^* \mp \frac{g(\eta^*)}{f'(\eta^*)}\,.
\end{align}
The ground state splitting is then (to linear order in $g$),
\begin{align}
\delta &= -2 L \frac{g(\eta^*)}{f'(\eta^*)}\,,
\end{align}
where the factor of $L$ comes from converting from $z$ to $\eta$ as in (\ref{fpmdef}). 

To simplify this further, we use the fact that $f(\eta) \approx 1 - 1/\eta$ so that $\eta^* \approx 1$. Therefore, to leading order, we have $f'(\eta^*) = 1$ and hence $\delta = -2 L g(1)$. Substituting into (\ref{gform}) we derive 
\begin{align}
\delta = 2\sqrt{\frac{L^{3+\alpha}}{\lambda}} \int_{0}^\infty \D{t}e^{-\sqrt{\frac{L^{1+\alpha}}{\lambda}} t} \ \<\Downarrow| \cos(P t) |\Uparrow\> \,.
\label{deltaformtoy}
\end{align}
Eq.~(\ref{deltaformtoy}) gives a semi-explicit formula for the ground state splitting $\delta$ and is the main result of our calculation. 

\subsection{Examples}
We now evaluate the ground state splitting $\delta$ (\ref{deltaformtoy}) for several different choices of $\mathcal{O}_j$. These different cases show that the scaling of $\delta$ can be sensitive to the precise form of the $\mathcal{O}_j$ operators. 

\subsubsection*{Example 1: $\mathcal{O}_j=\sigma_j^x$}
We focus on the case of even $L$ since otherwise the splitting $\delta = 0$ due to Kramers theorem. For even $L$, we have 
\begin{align}
    \<\Downarrow| \cos(Pt) |\Uparrow\> &= \frac{1}{2} \<\Downarrow| (e^{iPt} + e^{-iPt}) |\Uparrow\> \nonumber \\
    &= \frac{1}{2} \<\Downarrow| \prod_{k=1}^L (\cos t + i \sigma^x_k \sin t) |\Uparrow\> + c.c \nonumber \\
    &= (-1)^{L/2} \sin^L t\,.
    \label{cospt1}
\end{align}
Substituting this expression into (\ref{deltaformtoy}), we see that the integrand reaches its maximum magnitude at $t_* = \pi/2$ for $L$ sufficiently large. One can check that the integral is dominated by the neighborhood around this maximum, $t_* = \pi/2$ (for large $L$), so we can approximate $\sin^L t \approx e^{- L(t-\pi/2)^2/2}$. Evaluating the resulting integral and taking the logarithm, we obtain 
\begin{equation}
    \log |\delta|= -L^{\frac{1+\alpha}{2}} \frac{\pi}{2\sqrt{\lambda}} \,.
\end{equation}
to leading order in $L$. Thus, in this case, the toy model gives the same stretched exponential scaling as the all-to-all Ising spin chain (\ref{ising}).

\subsubsection*{Example 2: $\mathcal{O}_j=\sigma_j^x\sigma_{j+1}^x$}
Again we focus on even $L$ since otherwise $\delta = 0$ due to Kramers theorem. In this case we have
\begin{align}
    \<\Downarrow| e^{iPt} |\Uparrow\> &= \<\Downarrow| \prod_{k=1}^L (\cos t + i \sigma^x_k \sigma^x_{k+1} \sin t) |\Uparrow\> \nonumber \\
    &= 2 i^{L/2} \cos^{L/2} t \sin^{L/2} t\,,
\end{align}
so that
\begin{equation}
\<\Downarrow| \cos(Pt) |\Uparrow\> = i^{L/2} \cos^{L/2} t \sin^{L/2} t + c.c  \,.
\label{cospt2}
\end{equation}
Notice that if $L$ is not a multiple of $4$, then the above matrix element vanishes, and hence $\delta = 0$. 

Focusing on the case where $L$ is a multiple of $4$, we substitute the above expression into (\ref{deltaformtoy}) and observe that the integrand reaches its maximum magnitude at $t_* = \pi/4$. Again one can check that the integral is dominated by the neighborhood around $t = \pi/4$ in the large $L$ limit. Using the approximation $\cos^{L/2} t \sin^{L/2} t \approx \frac{1}{2^{L/2}} e^{-L(t-\pi/4)^2}$, which is valid in this neighborhood, and evaluating the resulting integral, we find that
\begin{align}
    \log |\delta| = - \frac{L \log 2}{2}\,,
\end{align}
to leading order in $L$. Evidently in this case, the splitting scales \emph{exponentially} in the system size, rather than as a stretched exponential. This exponential scaling of $\delta$ can be traced to the fact that the matrix element $|\<\Downarrow | \cos(Pt) |\Uparrow\>|$ (\ref{cospt2}) is exponentially small in $L$ for all $t$. This should be contrasted with the previous example where the matrix element $|\<\Downarrow | \cos(Pt) |\Uparrow\>|$ (\ref{cospt1}) is of order $1$ for some choices of $t$.

\subsubsection*{Example 3: $\mathcal{O}_j=\sigma_j^x+\gamma\sigma_j^x\sigma_{j+1}^x$}
Focusing on even $L$ again, we have
\begin{align}
    \<\Downarrow| e^{iPt} |\Uparrow\> &=  \<\Downarrow| \prod_{k=1}^L e^{it \sigma^x_k + i \gamma t \sigma^x_k \sigma^x_{k+1}} |\Uparrow\> \nonumber \\ 
    &= \frac{1}{2^{L}} \tr \left(\prod_{k=1}^L e^{it \sigma^x_k + i \gamma t\sigma^x_k \sigma^x_{k+1}} \prod_{k=1}^L \sigma^x_k \right) \,,
\end{align}
where the second equality follows from the fact that the only terms that contribute to the matrix element are those that have a single $\sigma^x_k$ on every site $k$.
Evaluating the resulting trace using a transfer matrix approach, one finds that
\begin{align}
\<\Downarrow| \cos(Pt) |\Uparrow\> &= \frac{1}{2^{L+1}} \tr T^L + c.c\,,
\end{align}
where $T$ is the $2 \times 2$ transfer matrix
\begin{align}
    T = \bpm e^{i t} e^{i\gamma t}  & i e^{-i \gamma t} \\ i e^{- i \gamma t} & -e^{-it} e^{i \gamma t} \epm\,.
\end{align}
The eigenvalues of $T$ are 
\begin{align}
\lambda_\pm = i e^{i \gamma t} [\sin t \pm (e^{-4 i \gamma t} - \cos^2 t)^{1/2}]\,,
\end{align}
so that
\begin{align}
\<\Downarrow| \cos(Pt) |\Uparrow\> &= \frac{1}{2^{L+1}} ( \lambda_+^L + \lambda_-^L)+ c.c\,.
\end{align}
The next step is to substitute this expression into (\ref{deltaformtoy}) and analyze the resulting integral. To proceed further, we specialize the case where
\begin{align}
\gamma = p/q 
\end{align}
is rational, and furthermore the denominator $q$ is odd, as this makes the analysis simple. In this case, one can check that the integrand of (\ref{deltaformtoy}) reaches its maximum magnitude at
$t_* = q \pi/2$. Also, one can check that the integral is dominated by the neighborhood around $t = t_*$: this follows from the fact that $|\lambda_{\text{max}}|$ reaches its maximum at $ t= t_*$ and also $\frac{d}{dt} \lambda_{\text{max}} |_{t= t^*} = 0$ where $\lambda_{\text{max}}$ denotes the eigenvalue $\lambda_\pm$ with the largest magnitude. Approximating $\<\Downarrow| \cos(Pt) |\Uparrow\> \sim c_1 e^{-c_2 L(t-t_*)^2}$ where $c_1, c_2$ are constants that depend on $p, q$ and evaluating the resulting integral, we obtain 
\begin{equation}
    \log |\delta|= -L^{\frac{1+\alpha}{2}} \frac{q\pi}{2\sqrt{\lambda}} \,.
\end{equation}
We conclude that the splitting scales like a stretched exponential as in the first example, but the constant in the exponent depends on the denominator $q$. 

\begin{table}
\begin{center}
    \begin{tabular}{|c|c|c|}
    \hline
        Model & \ Hamiltonian \ & \ Ground state splitting $\delta$ \ \\
        \hline
        \makecell{Summable Ising spin chain: \\ $\sum_r |f(r)| \leq c$} & (\ref{H}) & $\exp(-CL)$ \\
        \hline
       \makecell{All-to-all Ising spin chain: \\ $f(r) \sim 1/L^\alpha$, $0 < \alpha<1$} & (\ref{ising}) &$\exp(-C L^{\frac{1+\alpha}{2}})$ \\
       \hline
       \makecell{Quantum rotor chain: \\
       $f(r) \sim 1/|r|^\alpha$, $0 < \alpha<1$} & (\ref{confinedh}) &$\exp(-CL^{\frac{1+\alpha}{2}})$ \\
       \hline
       \makecell{All-to-all toy model: \\
       $f(r) \sim 1/L^\alpha$, $0 < \alpha < 1$} & (\ref{toymodel}) & $\exp(-CL^{\frac{1+\alpha}{2}})$\\
       \hline
    \end{tabular}
    \end{center}
    \caption{Scaling of the ground state energy splitting $\delta$ with system size $L$ for different variants of the Ising spin chain (\ref{H}). Here, $C$ denotes an $L$-independent constant.}
    \label{table}
\end{table}

\section{Discussion}
\label{sec:disc}

In summary, we have argued that the Ising spin chain (\ref{H}) with power-law interactions $f(r) \sim 1/r^\alpha$ is a useful toy model for studying the effect of long-range interactions on topological ground state degeneracy. We have particularly focused on the case where $\alpha < 1$ as these power-law interactions violate the summability condition (\ref{summability0}) and are therefore outside the regime of applicability of the results of Ref.~\onlinecite{lapa21stability}. While we were not able to analyze the power-law Ising spin chain directly for $\alpha < 1$, we successfully computed the ground state splitting for several closely related 1D models with long-range two-body interactions. First, we studied the Ising spin chain (\ref{ising}) with all-to-all interactions of the form $f(r) \sim 1/L^\alpha$. Next we considered a quantum rotor chain (\ref{confinedh}) with power-law interactions of the form $f(r) \sim 1/r^\alpha$. Finally, we studied a toy model (\ref{toymodel}) with all-to-all interactions of the form $f(r) \sim 1/L^\alpha$. In all of these cases, we found that the splitting $\delta$ scales like a stretched exponential $\delta \sim \exp[-(\text{const.}) L^{\frac{1+\alpha}{2}}]$ for $0 < \alpha < 1$ (though for the toy model (\ref{toymodel}) we also obtained other types of scaling, depending on the operators $\mathcal{O}_j$ that appear in the all-to-all interaction). These results are summarized in Table \ref{table}. 

Based on these results, we conjecture that the Ising spin chain (\ref{H}) with $f(r) \sim 1/r^\alpha$ also exhibits the same stretched exponential splitting. That is, for $0 < \alpha < 1$, the splitting scales as $\delta \sim \exp[-(\text{const.}) L^{\frac{1+\alpha}{2}}]$ . At the same time, the results of Ref.~\onlinecite{lapa21stability} guarantee that if $\alpha > 1$, then the splitting of the Ising spin chain model with $f(r) \sim 1/r^\alpha$ scales \emph{exponentially} with system size: $\delta \sim \exp[-(\text{const.}) L]$. Putting this together, our expectation for the marginal case $\alpha = 1$ is that the ground state splitting scales exponentially with system size $L$, possibly with logarithmic corrections. It would be interesting to test these conjectures numerically. 

It would also be interesting to investigate the scaling of the ground state splitting from a continuum field theory perspective. Perhaps such an approach could reveal whether the stretched exponential scaling that occurs in our models is a universal property or requires fine-tuning. 

 Another important direction for future work would be to study \emph{two-dimensional} (2D) systems with long range interactions. In particular it would be interesting to investigate a toy version of a fractional quantum Hall state with $1/r$ Coulomb interactions. A good starting point is the charge conserving toric code model discussed in Sec.~III~A of Ref.~\onlinecite{levin2011exactly} -- an exactly solvable gapped 2D lattice model with $U(1)$ charge conservation and fractionally charged anyon excitations. For this model, the key question would be to understand how the ground state splitting scales with system size if we add two types of perturbations, namely (1) weak $1/r$ density-density interactions and (2) a weak nearest neighbor charge hopping term. (The motivation for including the second perturbation is that it provides a finite dispersion to charged excitations, which are otherwise completely localized). It is plausible that the ground state splitting of this perturbed model is governed by the same physics as fractional quantum Hall states with Coulomb interactions.

Looking forward, if the stretched exponential ground state splitting reported here proves to be a general feature of topological qubits with slowly decaying power-law interactions, this would be a promising result for quantum information applications. It would indicate that such qubits remain largely robust in the presence of these interactions, with error rates that can still be efficiently suppressed -- namely, by a stretched exponential factor -- by increasing system size.
\acknowledgments 

We thank Mohsin Iqbal and Andrew Potter for comments on the manuscript. This work was supported by the Kadanoff Center for Theoretical Physics (E.G.), a Simons Investigator grant (M.L.) and by the
Simons Collaboration on Ultra-Quantum Matter (E.G. and M.L.), which is a grant from the Simons Foundation (651442).

\appendix

\section{Path minimizing $\Re(\mathcal{S}_{d}^{\rm tot})$ }\label{holder}
In this Appendix, we show that the constant path $\theta(\tau) = 0$ minimizes $\Re(\mathcal{S}_d[\theta])$ (\ref{SnSd}) and therefore also minimizes $\Re(\mathcal{S}_{d}^{\rm tot})$.
Our starting point is the H\"older inequality for matrices. Define the matrix p-norm $||X||_p$ by
\begin{equation}
    ||X||_p=(\tr(|X|^p))^{1/p}\,,
\end{equation}
with $|X|=\sqrt{X^\dagger X}$. The H\"older inequality states that for any $p,q,r\geq 1$ with $p^{-1}+q^{-1}=r^{-1}$ and any two matrices $X,Y$ (see section IV~2 of Ref.~\onlinecite{bhatia2013matrix})
\begin{equation}
    ||X Y||_r\leq ||X||_p ||Y||_q\,.
    \label{holderineq}
\end{equation} 
In what follows, we will need the following corollary of (\ref{holderineq}):
\begin{align}
    |\tr(U X_1 \cdots X_m)|&\leq \prod_{i=1}^m ||X_i||_m\
    \label{holdercorollary}
\end{align}
for any matrices $X_1,..., X_m$ and any unitary $U$. This corollary can be derived by noting that
\begin{align*}
    |\tr(U X_1 \cdots X_m)|&\leq \tr |X_1 \cdots X_m| \nonumber \\
    &=||X_1 \cdots X_m||_1\\
    &\leq \prod_{i=1}^m ||X_i||_m\,,
\end{align*}
where the first line follows from the properties of the polar decomposition of matrices, the second line follows from the definition of $||\cdot||_1$, and the last line follows from applying the H\"older inequality repeatedly.

To apply (\ref{holdercorollary}) in our case, we first rewrite the action $\mathcal{S}_d$ (\ref{SnSd}) using the Trotter expansion: 
\begin{equation}
    e^{-\mathcal{S}_d[\theta]}=\underset{m\to\infty}{\lim}\, \tr\left[ e^{-i M^x \theta(\beta)}\prod_{k=1}^m e^{-\frac{\beta}{m} W(\theta_k)}\right]\,,
\end{equation}
with $\theta_k=\theta(k \tfrac{\beta}{m})$. Fixing $m$ and setting
\begin{align}
    U = e^{-i M^x \theta(\beta)}, \quad X_k = e^{-\frac{\beta}{m} W(\theta_k)},
\end{align}
we have $\tr(|X_k|^m)=\tr e^{-\beta H_0}$, so that \eqref{holdercorollary} yields
\begin{align}
\left| \tr\left[ e^{-i M^x \theta(\beta)}\prod_{k=1}^m e^{-\frac{\beta}{m} W(\theta_k)}\right]\right| &\leq \prod_{k=1}^m [\tr e^{-\beta H_0}]^{1/m} \nonumber \\
&=\tr e^{-\beta H_0}\,.
\end{align}
Hence
\begin{equation}
    \left| e^{-\mathcal{S}_d[\theta]} \right|\leq \tr e^{-\beta H_0}\,.
\end{equation}
Since this inequality is saturated for $\theta=0$, we conclude that the constant path $\theta(\tau)=0$ minimizes $\Re(\mathcal{S}_d[\theta])$, as we wished to show.

\section{Paths minimizing $\Re(\mathcal{S}_{n}^{\rm tot})$ \label{saddlepointapp}}
In this Appendix, we argue that the paths $\bar{\theta}_n$ that minimize $\Re(\mathcal{S}_{n}^{\rm tot})$ have an approximate step-like structure with $\bar{\theta}_n(\tau) \approx 0$ for $\tau < \tau^*$ and $\bar{\theta}_n(\tau) \approx \pi$ for $\tau > \tau^*$ with the transition occurring over a time scale $\Delta \tau$ with $\Delta \tau \rightarrow 0$ as $L \rightarrow \infty$.

As in Appendix~\ref{holder}, our argument is based on the inequality (\ref{holdercorollary}). To apply this inequality, we first rewrite the action $\mathcal{S}_n$ (\ref{expsaz}) using the Trotter expansion:
\begin{align}
        e^{-\mathcal{S}_n[\theta]}=\underset{m\to\infty}{\lim}\, \tr\left[ S e^{-i M^x \theta(\beta)}\prod_{k=1}^m e^{-\frac{\beta}{m} W(\theta_k)}\right]\,,
\end{align}
with $\theta_k=\theta(k \tfrac{\beta}{m})$. Next we fix $m$ and apply (\ref{holdercorollary}) with
\begin{align}
    U = S e^{-i M^x \theta(\beta)}, \quad X_k = e^{-\frac{\beta}{m} W(\theta_k)}
\end{align}
to deduce that
\begin{align}
\left \|\tr\left[ S e^{-i M^x \theta(\beta)}\prod_{k=1}^m e^{-\frac{\beta}{m} W(\theta_k)}\right] \right \| \leq \tr e^{-\beta H_0}
\end{align}
We conclude that
\begin{equation}
\left| e^{- \mathcal{S}_n[\theta]} \right| \leq \tr e^{-\beta H_0}
\label{numineq}
\end{equation}
similarly to Appendix~\ref{holder}.

Next, we analyze the conditions under which the inequality (\ref{numineq}) is saturated. Recall that the usual H\"older inequality (\ref{holderineq}) is saturated if and only if $|X|^p \propto | Y^\dagger|^q$.  Therefore, if we consider the inequality (\ref{holdercorollary}) in the special case where the $X_k$ are Hermitian and invertible, we see that (\ref{holdercorollary}) is saturated if and only if $|X_i|^m \propto |X_j|^m$ for all $i, j$ and $U \propto |X_1 \cdots X_m| (X_1 \cdots X_m)^{-1} $. Applying the latter result to our derivation of (\ref{numineq}) where $X_k = e^{-\frac{\beta}{m} W(\theta_k)}$, the first condition reduces to $W(\theta_i) = W(\theta_j)$ or equivalently, $\theta_i = \theta_j \pmod{\pi}$. Likewise, the second condition reduces to $U \propto \mathbbm{1}$ or equivalently $\theta(\beta) = \pi \pmod{2\pi}$. 
Putting this all together, we conclude that the inequality (\ref{numineq}) is saturated if and only if (a) $\theta(\tau)$ is piecewise constant with jumps that are multiples of $\pi$, and (a) $\theta(\beta)$ is an odd multiple of $\pi$.

The above conditions (a) and (b) determine the paths that maximize $\left| e^{- \mathcal{S}_n[\theta]} \right|$, or equivalently, minimize $\Re(\mathcal{S}_n[\theta])$.
Paths that violate these conditions will have a strictly larger value of $\Re(\mathcal{S}_n[\theta])$. How much larger? Consider a path that deviates from the above piecewise constant behavior for some time interval $\Delta \tau$. We expect that paths of this type will have an action that is larger than the minimum possible value by a factor of the form
\begin{align}
    \Re(\mathcal{S}_n[\theta]) \geq -\log(\tr e^{-\beta H_0}) + (\text{const.}) L \Delta \tau
    \label{approxpwc1}
\end{align}
Here, the second term is expected on general grounds since the action is expected to scale extensively in the system size $L$. (One can also derive this term more explicitly from a microscopic calculation. The basic idea is to use a Jordan-Wigner transformation to relate $S_n[\theta]$ to a free fermion expression. One can then express $S_n[\theta]$ as a sum of terms, one for each free fermion mode, labeled by wave vectors $k$. Each $k$ term will be increased from the minimum possible value by a constant amount, and therefore the product over all $k$ is suppressed by an exponential factor since the number of $k$'s scales linearly with $L$).

Since the kinetic term is always positive, the total action obeys a similar bound:
\begin{align}
    \Re(\mathcal{S}_n^{\rm tot}[\theta]) \geq -\log(\tr e^{-\beta H_0}) + (\text{const.}) L \Delta \tau
    \label{approxpwc2}
\end{align}
To complete the argument, we now compare (\ref{approxpwc2}) to the action $\mathcal{S}_{n0}$ for the path $\bar{\theta}_n$ (\ref{thetanform}) found in Section~\ref{minimalpath}: in particular, from (\ref{sn0}) we have
\begin{align}
\mathcal{S}_{n0} = -\log(\tr e^{-\beta H_0}) + (\text{const.}) L^{\frac{1+\alpha}{2}}
\end{align}
Evidently, the only way that a path can be competitive with $\bar{\theta}_n$ (\ref{thetanform}) in terms of minimizing $\Re(\mathcal{S}_n^{\rm tot})$ is if $\Delta \tau \lesssim L^{\frac{\alpha-1}{2}}$. In other words, any candidate path must be piecewise constant, outside of a short time interval $\Delta \tau$ that approaches $0$ as $L \rightarrow \infty$.
Similar reasoning implies that $\theta(\beta) = \pi \pmod{2\pi}$ up to a correction that vanishes as $L \rightarrow \infty$. Putting this all together, this establishes that any candidate path for minimizing $\Re(\mathcal{S}_n^{\rm tot})$ must have the approximate step-like structure that we claimed above.

\section{Scaling of the ratio of determinants in Eq.~(\ref{ratiodetermscaling})}
\label{hessiannegligible}

In this Appendix, we derive Eq.~(\ref{ratiodetermscaling}). That is, we derive the scaling behavior
\begin{align}
      \frac{\det \mathcal{H}_d}{\det' \mathcal{H}_n} \propto L
      \label{ratiodetapp}
\end{align}
in the limit $L \rightarrow \infty$. Here the Hessian operator $\mathcal{H}_d$ is defined by expanding the action $\mathcal{S}^{\mathrm{tot}}_d$ to quadratic order about its saddle point at $\theta = 0$:
\begin{align}
    \mathcal{S}_d^{\mathrm{tot}}[\delta \theta] = \mathcal{S}_{d0} + \frac{1}{2} \delta \theta \cdot \mathcal{H}_d \cdot \delta \theta
\end{align}
Likewise, $\mathcal{H}_n$ is given by expanding $\mathcal{S}^{\mathrm{tot}}_n$ about its saddle point at $\theta = \bar{\theta}_n$:
\begin{align}
    \mathcal{S}_n^{\mathrm{tot}}[\bar{\theta}_n+\delta \theta] = \mathcal{S}_{n0} + \frac{1}{2} \delta \theta \cdot \mathcal{H}_n \cdot \delta \theta
\end{align}

Our approach is somewhat indirect: instead of directly evaluating the ratio of determinants in (\ref{ratiodetapp}), we will instead evaluate another, closely related, ratio of determinants. The latter ratio of determinants is defined as follows. Consider the actions $\mathcal{S}_d^{\mathrm{tot}}[\varphi]$ and $\mathcal{S}_n^{\mathrm{tot}}[\varphi]$ expressed in terms of our \emph{original} variables $\varphi = \dot{\theta}$. Consider the corresponding Hessians, $\mathcal{H}^{(\varphi)}_d$ and $\mathcal{H}^{(\varphi)}_n$, defined by
\begin{align}
    \mathcal{S}_d^{\mathrm{tot}}[\delta \varphi] &= \mathcal{S}_{d0} + \frac{1}{2} \delta \varphi \cdot \mathcal{H}^{(\varphi)}_d \cdot \delta \varphi \nonumber \\
    \mathcal{S}_n^{\mathrm{tot}}[\bar{\varphi}_n+\delta \varphi] &= \mathcal{S}_{n0} + \frac{1}{2} \delta \varphi \cdot \mathcal{H}^{(\varphi)}_n\cdot \delta \varphi   
\end{align}
where the saddle point $\bar{\varphi}_n$ is defined by $\bar{\varphi}_n = \dot{\bar{\theta}}_n$. We will be interested in the ratio of determinants $\frac{\det \mathcal{H}^{(\varphi)}_d}{\det' \mathcal{H}^{(\varphi)}_n}$. We will argue that this ratio scales as
\begin{align}
\frac{\det \mathcal{H}^{(\varphi)}_d}{\det' \mathcal{H}^{(\varphi)}_n} \propto L^\alpha
\label{ratiodetphiscal}
\end{align}
Once we establish (\ref{ratiodetphiscal}), we will be done. Indeed, as we explain below, one can use a change of variables to show that
\begin{align}
    \frac{\det \mathcal{H}_d}{\det' \mathcal{H}_n} \propto \frac{\det \mathcal{H}^{(\varphi)}_d}{\det' \mathcal{H}^{(\varphi)}_n} L^{1-\alpha}
\label{changevaridHnHd}
\end{align}
The scaling in Eq.~(\ref{ratiodetapp}) then follows immediately from Eq.~(\ref{ratiodetphiscal}). 

We now derive the relation (\ref{changevaridHnHd}). Our starting point is the following linear algebra identity. Let $W$ be an invertible matrix and let $A, B$ be Hermitian matrices such that $B$ has a single zero eigenvalue. We claim that
\begin{align}
\frac{\det(W^\dagger A W)}{\det'( W^\dagger B W)} = \frac{\det A}{\det' B} 
\frac{\<\psi|W^\dagger W|\psi\>}{\<\psi| \psi\>}
\label{changevarid}
\end{align}
where $|\psi\>$ denotes the (unique) eigenvector of $W^\dagger B W$ with eigenvalue $0$, and $W |\psi\>$ is the corresponding zero mode of $B$. To derive (\ref{changevarid}), notice that
\begin{align}
\frac{\det A}{\det' B} 
&= \epsilon \lim_{\epsilon \rightarrow 0} \frac{\det A}{\det(B + \epsilon \mathbbm{1})} \nonumber \\
&=\epsilon \lim_{\epsilon \rightarrow 0} \frac{\det(W^\dagger A W)}{\det(W^\dagger B W + \epsilon W^\dagger W)} \nonumber \\
&= \frac{\det(W^\dagger A W)}{\det'(W^\dagger B W)}  \frac{\<\psi| \psi\>}{\<\psi|W^\dagger W|\psi\>}
\end{align}
where the first equality follows from the definition of $\det'(B)$, the second equality follows from the multiplicative property of the determinant, and the last equality follows from the (first order) perturbative expansion for the lowest eigenvalue of $W^\dagger B W + \epsilon W^\dagger W$.

To derive (\ref{changevaridHnHd}) from (\ref{changevarid}), let $A = \mathcal{H}_d^{(\varphi)}$ and $B = \mathcal{H}_n^{(\varphi)}$ and $W = \frac{d}{d\tau}$. Then $W^\dagger A W = \mathcal{H}_d$ and $W^\dagger B W = \mathcal{H}_n$ and $|\psi\> = \dot{\bar{\theta}}_n$, so (\ref{changevarid}) gives
\begin{align}
    \frac{\det \mathcal{H}_d}{\det' \mathcal{H}_n} = \frac{\det \mathcal{H}^{(\varphi)}_d}{\det' \mathcal{H}^{(\varphi)}_n} \frac{\int_0^\beta d\tau \ \ddot{\bar{\theta}}_n^2}{\int_0^\beta d\tau \ \dot{\bar{\theta}}_n^2}
\end{align}
Eq.~(\ref{changevaridHnHd}) now follows immediately since the ratio of integrals on the right hand side scales as
\begin{align}
    \frac{\int_0^\beta d\tau \ \ddot{\bar{\theta}}_n^2}{\int_0^\beta d\tau \ \dot{\bar{\theta}}_n^2} \propto L^{1-\alpha}
\end{align}
due to the fact that $\bar{\theta}_n$ (\ref{thetanform}) has a characteristic time scale of $\Delta \tau \propto L^{\frac{\alpha-1}{2}}$. 

At this point, all that remains is to show (\ref{ratiodetphiscal}). We begin by studying the two Hessians $\mathcal{H}^{(\varphi)}_d$ and $\mathcal{H}^{(\varphi)}_n$ that appear in (\ref{ratiodetphiscal}). Using (\ref{SnSdphi}), we can see that the Hessian $\mathcal{H}^{(\varphi)}_d$ is given by
\begin{align}
\mathcal{H}^{(\varphi)}_d = \frac{L^\alpha}{2 \lambda} \mathbbm{1} + V_d
\label{hd0vd}
\end{align}
where $V_d$ is an integral operator with kernel
\begin{align}
V_d(\tau_1, \tau_2) = \<M_x(\tau_1) M_x(\tau_2)\>_0 - \<M_x(\tau_1)\>_0 \<M_x(\tau_2)\>_0
\end{align}
and where the above correlation functions are defined by
\begin{align}
\<M_x(\tau_1) &M_x(\tau_2)\>_0 = \nonumber \\
&\frac{\tr\left(e^{-(\beta - \tau_1) H_0 } M_x e^{-(\tau_1 - \tau_2) H_0} M_x e^{-\tau_2 H_0}\right)}{\tr(e^{-\beta H_0})}
\end{align}
(for $\tau_1 \geq \tau_2$) and
\begin{align}
\<M_x(\tau)\>_0 = \frac{\tr\left(e^{-(\beta - \tau) H_0 } M_x  e^{-\tau H_0} \right)}{\tr(e^{-\beta H_0})}    
\end{align}
Similarly, the Hessian $\mathcal{H}^{(\varphi)}_n$ is given by
\begin{align}
\mathcal{H}^{(\varphi)}_n = \frac{L^\alpha}{2 \lambda} \mathbbm{1} + V_n
\end{align}
where $V_n$ is an integral operator with kernel
\begin{align}
V_n(\tau_1, \tau_2) &= \< M_x(\tau_1) M_x(\tau_2)\>_{\bar{\varphi}_n} \nonumber \\
&- \<M_x(\tau_1)\>_{\bar{\varphi}_n} \<M_x(\tau_2)\>_{\bar{\varphi}_n}
\label{vncorr}
\end{align}
where 
\begin{align}
\<M_x(\tau_1)&M_x(\tau_2)\>_{\bar{\varphi}_n} = \nonumber \\
&\frac{\tr[U(\beta,\tau_1)  M_x U(\tau_1, \tau_2) M_x U(\tau_2, 0) ]}{\tr[U(\beta,0)]}
\end{align}
(for $\tau_1 \geq \tau_2$) and
\begin{align}
\<M_x(\tau)\>_{\bar{\varphi}_n} = \frac{\tr[ U(\beta,\tau) M_x U(\tau, 0)]}{\tr[U(\beta,0)]}    
\end{align}
with
\begin{align}
U(\tau_1, \tau_2) = \mathcal{T} \exp \left(-\int_{\tau_2}^{\tau_1} d \tau W(\bar{\theta}_n(\tau)) \right)
\end{align}

In the case of $V_d$, the correlation function can be evaluated easily, giving
\begin{align}
V_d(\tau_1, \tau_2) = \frac{L}{4} \frac{\cosh^2(\beta - 2 |\tau_1 - \tau_2|)}{\cosh^2 \beta}    
\end{align}
in the limit of large $L$. We can also find an explicit formula for the eigenvalues and eigenvectors of $V_d$. In particular, since $V_d$ is translationally invariant (i.e. $V_d(\tau_1, \tau_2) = V_d(\tau_1 + \tau, \tau_2 + \tau)$) the eigenvectors of $V_d$ are of the form $\psi_k(\tau) = e^{i \omega_k \tau}$ where $\omega_k = 2\pi k/\beta$ and $k$ is an integer. The corresponding eigenvalues $v_{k}$ are given by
\begin{align}
    v_{k} = \frac{L}{\cosh^2 \beta}\left(\frac{\sinh 2\beta}{16 + \omega_k^2} + \frac{\beta}{8} \delta_{k,0} \right)
    \label{lambdak}
\end{align}

Using (\ref{lambdak}), we can write down an explicit formula for the eigenvalues of $\mathcal{H}_d^{(\varphi)}$, which we will denote by $h_{k}$. In particular, combining (\ref{lambdak}) with the constant term in (\ref{hd0vd}) we obtain
\begin{align}
    h_{k} = \frac{L^\alpha}{2 \lambda} + \frac{L}{\cosh^2 \beta}\left(\frac{\sinh 2\beta}{16 + \omega_k^2} + \frac{\beta}{8} \delta_{k,0} \right)
\end{align}
From this formula, we see that the eigenvalues of $\mathcal{H}_d^{(\varphi)}$ range from a minimum of $L^\alpha/2\lambda$ (which occurs in the limit $k \rightarrow \infty$) to a maximum of order $L$ (which occurs when $k = 0$).

Unfortunately, we do not have a similarly explicit formula for the eigenvalues of $\mathcal{H}_n^{(\varphi)}$. The problem is that the kernel $V_n(\tau_1, \tau_2)$ is not translationally invariant since the path $\bar{\varphi}_n$ explicitly breaks translational symmetry. This lack of translational invariance makes it difficult to diagonalize $V_n$ analytically. There are several ways to proceed. One approach is to numerically diagonalize $V_n$ and then numerically compute the ratio of determinants. Alternatively, one can try to estimate the scaling of the ratio of determinants using simple arguments. We will follow the second approach here. 

Let us consider the ratio of determinants $\frac{\det \mathcal{H}^{(\varphi)}_d}{\det' \mathcal{H}^{(\varphi)}_n}$ . To begin, it is helpful to regularize this ratio by discretizing (imaginary) time $\tau$. Such a discretization turns both $\mathcal{H}_d^{(\varphi)}$ and $\mathcal{H}_n^{(\varphi)}$ into finite dimensional matrices so that the determinants are well-defined and finite. 

Imagine fixing a discretization, say with $m$ points, so that $\mathcal{H}_d^{(\varphi)}$ and $\mathcal{H}_n^{(\varphi)}$ are each $m \times m$ matrices. Now imagine pairing off the eigenvalues of the two matrices, starting with the largest eigenvalues of each matrix and then continuing in descending order. We expect that the ratio of each eigenvalue of $\mathcal{H}_d^{(\varphi)}$ and its partner in $\mathcal{H}_n^{(\varphi)}$ is of order $1$. Importantly, however, the smallest eigenvalue of $\mathcal{H}_d^{(\varphi)}$ has no partner since the corresponding eigenvalue of $\mathcal{H}_n^{(\varphi)}$ is $0$, which is not included in $\det' \mathcal{H}^{(\varphi)}_n$. This line of reasoning suggests that the ratio of determinants $\frac{\det \mathcal{H}^{(\varphi)}_d}{\det' \mathcal{H}^{(\varphi)}_n}$ should scale like the smallest eigenvalue of $\mathcal{H}_d^{(\varphi)}$ (since all the other eigenvalues can be paired off with each other, giving a ratio of order $1$). Therefore, since the smallest eigenvalue is approximately $L^\alpha/2 \lambda$, we deduce that the ratio of determinants scales as $L^\alpha$, as in (\ref{ratiodetphiscal}), as we wished to show.

\section{Computing the ratio of determinants \eqref{detratiorot} \label{apphess}}
In this Appendix, we compute the ratio of determinants in (\ref{detratiorot}). The first step in our calculation is to solve the eigenvalue problem for the two Hessian operators $\mathcal{H}^{(1)}$ and $\mathcal{H}^{(2)}$ that appear in (\ref{detratiorot}).  

\subsection{Eigenvalues of $\mathcal{H}^{(1)}$}
We start with the Hessian operator $\mathcal{H}^{(1)}$. This operator is defined by expanding $\mathcal{S}$ to quadratic order about $\bm{\theta}=-\pi/2$:
\begin{equation}
    \mathcal{H}^{(1)}_{ij}=-M_{ij}\partial_\tau^2+(2\delta_{ij}-\delta_{i,j+1}-\delta_{i,j-1})+\delta_{ij}\,.
\end{equation}
The eigenfunctions of $\mathcal{H}^{(1)}$ satisfy
\begin{equation}\label{diff}
     \mathcal{H}^{(1)}\delta \bm{\theta}=\eta \delta \bm{\theta}\,,
\end{equation}
where $\eta$ is the eigenvalue we wish to determine. In principle, our task is to find solutions to the above differential equation (\ref{diff}) for $\tau \in (-\infty, \infty)$ with the boundary condition that $\delta \bm{\theta}(\pm \infty) = 0$. However, in what follows we will regulate this eigenvalue problem by considering the equation on a finite interval $[-\beta/2, \beta/2]$ with the boundary condition $\delta \bm{\theta}(\pm\beta/2) = 0$. We will then take the limit $\beta \rightarrow \infty$ at the end of the calculation.

To proceed, notice that $\mathcal{H}^{(1)}$ is invariant under $\tau \rightarrow - \tau$, so we can choose $\delta \bm{\theta}(\tau)$ to be an even or odd function of $\tau$. For the even eigenstates, we use the ansatz
\begin{equation}
    \delta \bm{\theta}(\tau)= \vec{X}\cos (\Omega \tau) \,,
\end{equation}
where $\vec{X}$ is an $L$ component vector. The boundary conditions at $\tau = \pm \beta/2$ require that $\Omega=\frac{2\pi (n+1/2)}{\beta}$ for some integer $n \geq 0$. Substituting this ansatz into the differential equation \eqref{diff}, we deduce that $\vec{X}$ has to be an eigenvector of the matrix
\begin{equation}
    T_{ij}(u)=u^2 M_{ij}+(2\delta_{ij}-\delta_{i,j+1}-\delta_{i,j-1})
\end{equation}
at $u=\Omega$. Since $T(u)$ is a circulant matrix, it is diagonalized by a Fourier transform, and its eigenvalues are
\begin{equation}
    \tau_k(u)=u^2 m_k+4\sin^2(\pi k/L)\,,
\end{equation}
where $m_k$ are the eigenvalues of $M_{ij}$, i.e.
\begin{equation}
    \frac{1}{m_k}=1+2\lambda \sum_{r=0}^{L-1} \cos(2\pi k r/L)f(r)\,.
    \label{mkeq}
\end{equation}
and where $k = 0,1,...,L-1$. Hence the eigenvalues $\eta$ of $\mathcal{H}^{(1)}$ associated to even eigenstates are
\begin{equation}
    \tau_k(\tfrac{2\pi(n+1/2)}{\beta})+1\,,\qquad n\geq 0\,,\qquad k=0,1,...,L-1\,.
\label{eveneigh1}
\end{equation}

We now move on to the \emph{odd} eigenstates. In this case, we use the ansatz
\begin{equation}
    \delta \bm{\theta}(\tau)= \vec{X} \sin (\Omega \tau) \,,
\end{equation}
and the boundary conditions require that $\Omega=\frac{2\pi n}{\beta}$ for some integer $n \geq 1$. Again, $\vec{X}$ must be an eigenvector of $T(\Omega)$. Therefore following the same calculation as above, we find that the eigenvalues $\eta$ of $\mathcal{H}^{(1)}$ associated to odd eigenstates are
\begin{equation}
    \tau_k(\tfrac{2\pi n}{\beta})+1\,,\qquad n\geq 1\,,\qquad k=0,1,...,L-1\,.
    \label{oddeigh1}
\end{equation}

\subsection{Eigenvalues of $\mathcal{H}^{(2)}$}
Next we consider the Hessian operator $\mathcal{H}^{(2)}$. This operator is defined by expanding $\mathcal{S}$ to quadratic order around $\bm{\theta}=\bar{\bm{\theta}}$:
\begin{equation*}
    \mathcal{H}^{(2)}_{ij}=-M_{ij}\partial_\tau^2+(2\delta_{ij}-\delta_{i,j+1}-\delta_{i,j-1})+U''(\bar{\theta}(\tau)) \delta_{ij}\,.
\end{equation*}
Here we take $\bar{\theta}$ to be the instanton path $\bar{\theta}(\tau; \tau^*)$ given in (\ref{instantonrot}) with $\tau^* = 0$. Plugging in our specific choice of $U(\theta)$ \eqref{u} along with the formula for $\bar{\theta}$ (\ref{instantonrot}), we obtain
\begin{align}
    \mathcal{H}^{(2)}_{ij}&=-M_{ij}\partial_\tau^2+(2\delta_{ij}-\delta_{i,j+1}-\delta_{i,j-1})\nonumber \\ &+[1-2\sqrt{m_0}\delta (\tau) ]\delta_{ij}\,.
\end{align}

As in the previous section, our task is to find eigenfunctions of $\mathcal{H}^{(2)}$ satisfying
\begin{equation}\label{diff2}
     \mathcal{H}^{(2)}\delta \bm{\theta}=\eta \delta \bm{\theta}\,,
\end{equation}
Again, we impose the boundary condition that $\delta \bm{\theta}(\pm \beta/2) = 0$ with the idea of taking the limit $\beta \rightarrow \infty$ at the end of the calculation. 

As before, we can choose the eigenstates of $\mathcal{H}^{(2)}$ to be either even or odd functions of $\tau$. We start with the \emph{odd} eigenstates. In this case, the $\delta$ function term has no effect, so $\mathcal{H}^{(2)}$ has the same odd eigenvalues \eqref{oddeigh1} as $\mathcal{H}^{(1)}$.

As for even eigenstates, in this case the $\delta$ function imposes the matching condition 
\begin{equation}
    M \delta \bm{\theta}'(0^+)=-M \delta \bm{\theta}'(0^-)=-\sqrt{m_0}\delta \bm{\theta}(0)\,.
\end{equation}
This matching condition is satisfied by the ansatz
\begin{equation}
    \delta \bm{\theta}(\tau)=  \left(\cos (\Omega \tau)-\frac{ \sqrt{m_0}}{\Omega}\sin(\Omega |\tau|)M^{-1} \right) \vec{X} \,.
\label{h2eigfunct}
\end{equation}
Substituting this ansatz into the differential equation (\ref{diff2}), and using the fact that $M$ commutes with $T(u)$ (since they are both circulant matrices), we see that $\vec{X}$ is an eigenvector of $T(\Omega)$, just like in the previous cases. Finally, imposing the boundary condition $\delta {\theta}(\pm \beta/2)= 0$, we obtain the following equation on $\Omega$: 
\begin{equation}\label{betheeq}
    \tan \left(\frac{\Omega\beta}{2} \right)=\frac{\Omega m_k}{\sqrt{m_0}}\,,
\end{equation}
where $m_k$ is given in (\ref{mkeq}). Hence the eigenvalues of $\mathcal{H}^{(2)}$ associated with even eigenstates are
\begin{equation}
    \tau_k(\Omega)+1\,,\qquad k=0,1,...,L-1
    \label{eveneigh2}
\end{equation}
with $\Omega$ satisfying \eqref{betheeq}.

\subsection{Ratio of determinants}
To compute the ratio of determinants \eqref{detratiorot}, let us study the solutions $\Omega$ to the equations \eqref{betheeq} in the limit of large $\beta$. First consider solutions for which $\Omega$ is \emph{complex}. Without loss of generality, we can assume that $\Omega$ has a positive imaginary part, since $\Omega$ and $-\Omega$ correspond to the same eigenfunction (\ref{h2eigfunct}). If $\Omega$ has a positive imaginary part, then in the limit $\beta\to \infty$ one has $\tan(\Omega \beta/2)\to i$, and so
\begin{equation}
    \Omega=i\frac{\sqrt{m_0}}{m_k}
    \label{complexom}
\end{equation}
in the limit $\beta\to \infty$. We conclude that (\ref{complexom}) is the only complex solution to \eqref{betheeq} for each value of $k=0,1,...,L-1$. 

As for the real solutions to \eqref{betheeq}, they can be labeled by a positive integer $n \geq 1$ where $\Omega_n \in [2\pi n/\beta, 2\pi (n+1/2)/\beta]$. While there is no closed form expression for these solutions $\Omega_n$, one can make expansions in various limits. For our purposes, we will need to understand the behavior of $\Omega_n$ in the limit where $\beta$ is large and $n$ is of order $\beta$. In this regime, $\Omega_n$ can be approximated as
\begin{equation}\label{expansion}
    \Omega_n=\frac{2\pi(n+1/2)}{\beta}-\frac{2}{\beta}\arctan \left(\frac{\beta\sqrt{m_0}}{2\pi n m_k }\right)+\mathcal{O}(\beta^{-2})\ .
\end{equation}

We are now ready to compute the ratio of determinants in \eqref{detratiorot}, which we denote by $\Delta$:
\begin{align}
    \Delta = \left| \frac{\det \mathcal{H}^{(1)}}{\det' \mathcal{H}^{(2)}} \right|
\end{align}
Using the eigenvalues of $\mathcal{H}^{(1)}$ (\ref{eveneigh1}, \ref{oddeigh1}), we have
\begin{widetext}
\begin{align}
    \left| \det \mathcal{H}^{(1)} \right| = \prod_{k=0}^{L-1} ( \tau_k( \tfrac{\pi}{\beta}) + 1) \cdot \left[\prod_{k=0}^{L-1} \prod_{n \geq 1} ( \tau_k( \tfrac{2\pi (n+1/2)}{\beta}) + 1) ( \tau_k( \tfrac{2\pi n}{\beta}) + 1)  \right]
    \label{deth1}
\end{align}
Likewise, using the eigenvalues of $\mathcal{H}^{(2)}$ (\ref{eveneigh2}-\ref{expansion}),
we have
\begin{align}
        \left| \text{det}' \ \mathcal{H}^{(2)} \right| =  \prod_{k=1}^{L-1} ( \tau_k( i \tfrac{\sqrt{m_0}}{m_k}) + 1) \cdot \left[\prod_{k=0}^{L-1} \prod_{n \geq 1}  ( \tau_k(\Omega_n) + 1)( \tau_k( \tfrac{2\pi n}{\beta}) + 1)\right] 
        \label{deth2}
\end{align}
(Here, the reason that the product over $k$ in the first term doesn't include $k=0$ is that the $k=0$ term corresponds to the zero mode, which we are supposed to remove: that is, $\tau_k( i \tfrac{\omega \sqrt{m_0}}{m_k}) + 1 = 0$ for $k =0$). 

Taking the ratio of (\ref{deth1}) and (\ref{deth2}) gives:
\begin{equation}
    \Delta=(\tau_0(\tfrac{\pi}{\beta})+1)\prod^{L-1}_{\substack{k=1}}\frac{\tau_k(\tfrac{\pi}{\beta})+1}{\tau_k(i\tfrac{\sqrt{m_0}}{m_k })+1}\prod^{L-1}_{k=0}\prod_{n\geq 1}\frac{\tau_k(\tfrac{2\pi(n+1/2)}{\beta})+1}{\tau_k(\Omega_{n})+1}\,.
\end{equation}
We now take the limit $\beta\to\infty$. To that end, we write the second product as an exponential of a log and convert the sums over $n$ into integrals over $x=\frac{2\pi(n+1/2)}{\beta}$, obtaining
\begin{equation}
   \Delta= \prod_{k=1}^{L-1}\frac{\tau_k(0)+1}{\tau_k(i\tfrac{\sqrt{m_0}}{m_k })+1}\prod_{k=0}^{L-1}\exp\left(\frac{2}{\pi}\int_0^\infty \D{x} \frac{m_k x \arctan (\tfrac{\sqrt{m_0}}{m_k x})}{\tau_k(x)+1} \right)\,.
\end{equation}
Finally, using the identity
\begin{equation}
\int_0^\infty \frac{x \arctan(\tfrac{a}{ x})}{b^2+x^2}\D{x}=\frac{\pi}{2}\log\left(1+\frac{a}{b}\right)\,,
\end{equation}
we obtain the following exact value for the ratio of determinants $\Delta$:
\begin{equation}
    \Delta=2\prod_{k=1}^{L-1} \left(1-\frac{1}{\sqrt{1+4 \sin^2(\tfrac{\pi k}{L})}}\sqrt{\frac{m_0}{m_k}} \right)^{-1}\,.
\end{equation}
\end{widetext}

\end{document}